\documentclass{preprint}
\usepackage{dslabmath}

\title{A Generative Diffusion Model for Amorphous Materials}

\author
{Kai Yang,$^{1,2}$
Daniel Schwalbe-Koda$^{1,*}$\\
\vspace{1em} 
\normalfont{
\small
$^{1}$Department of Materials Science and Engineering, University of California, Los Angeles, CA, USA \\
$^{2}$Department of Civil and Environmental Engineering, University of California, Los Angeles, CA, USA\\
$^{*}$E-mail: \href{mailto:dskoda@ucla.edu}{dskoda@ucla.edu}
}
}

\usepackage[resetlabels]{multibib}
\newcites{Supp}{Supplementary References}
\newcites{Main}{References}

\newcommand{\supptext}{\ref{sec:stext}}
\newcommand{\methods}{\ref{sec:methods}}

\begin{document}

\maketitle
\thispagestyle{firstpagestyle} 

\begin{abstract}
Generative models show great promise for the inverse design of molecules and inorganic crystals, but remain largely ineffective within more complex structures such as amorphous materials.
Here, we present a diffusion model that reliably generates amorphous structures up to 1000 times faster than conventional simulations across processing conditions, compositions, and data sources.
Generated structures recovered the short- and medium-range order, sampling diversity, and macroscopic properties of silica glass, as validated by simulations and an information-theoretical strategy.
Conditional generation allowed sampling large structures at low cooling rates of 10$^{-2}$ K/ps to uncover a ductile-to-brittle transition and mesoporous silica structures.
Extension to metallic glassy systems accurately reproduced local structures and properties from both computational and experimental datasets, demonstrating how synthetic data can be generated from characterization results.
Our methods provide a roadmap for the design and simulation of amorphous materials previously inaccessible to computational methods.
\end{abstract}

\section{Introduction}

Solving the inverse design of materials is one of the most coveted goals of computational materials science,\cite{zunger2018inverse} but sampling the space of materials exhibiting outstanding structures and properties remains challenging.
In recent years, deep generative models provided an effective framework for sampling data points conditioned on labels.\cite{goodfellow2014generative}
Diffusion models, in particular, excel at learning the joint probability distributions of complex data and labels, denoising random inputs into high-fidelity samples such as images, text, and more.\cite{ho2020denoising,yang2023diffusion}
Within materials science, diffusion models have been shown to produce novel inorganic crystal structures conditioned on properties, symmetries, compositions, and other constraints, in an encouraging step for the inverse design of crystalline materials.\cite{xie2022crystal,zeni2025generative, guo2025ab, zhong2025practical}

Many technologically relevant solids, however, are not crystalline, but amorphous.
Polymers, metallic glasses, battery electrolytes, and phase-change memory alloys are examples of materials lacking the long-range order that makes inorganic crystalline systems comparatively simpler to represent and, in many cases, generate.\cite{schwalbe2020generative,liu2024amorphous}
Furthermore, the rugged, high-dimensional energy landscapes of amorphous structures demand very long molecular dynamics (MD) or Monte Carlo simulations to achieve adequate convergence.\cite{urata2024applications}
Thus, modeling synthesis-structure-property relationships that can enable the inverse design of amorphous materials remains challenging.
Several works have attempted to use generative models, including variational autoencoders, generative adversarial networks, or diffusion models, to produce amorphous configurations.\cite{comin2019deep,kilgour2020generating,chen2025physical, li2025conditional,xu2023generative,kwon2024Spectroscopy-guided}
However, comparisons between generated and simulated structures show that most models fail to either produce physically correct samples\cite{comin2019deep} or sample structures well outside the training set.\cite{chen2025physical,yong2024dismai}
Even when structures are reasonable, there is limited evidence that generated samples exhibit the correct macroscopic properties,\cite{kwon2024Spectroscopy-guided} severely limiting the applicability of generative models in computational studies (\supptext, Sec. \ref{sec:si:evaluating}).

\begin{figure}[tb!]
    \centering
    \includegraphics[width=\textwidth]{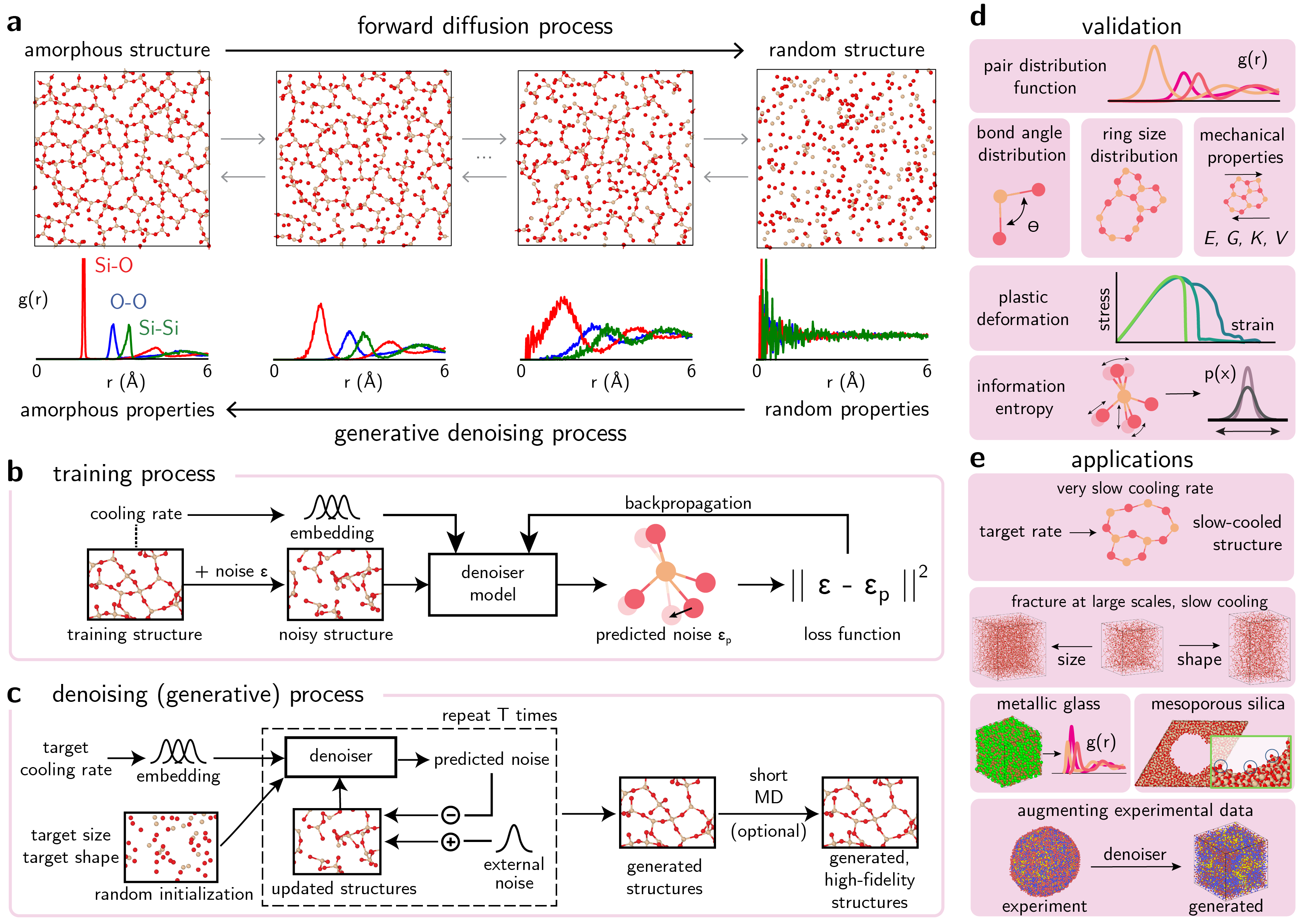}
    \caption{
    \textbf{Training and validating a generative diffusion model for amorphous materials.}
    \textbf{a}, Forward diffusion and reverse denoising process for amorphous materials.
    The structure and partial pair distribution functions (PDF) for SiO$_2$ are provided as an example.
    In the forward process, noise is progressively added to atomic positions until the structure becomes random.
    In the reverse (generative) direction, the model gradually denoises the positions of randomly sampled positions and creates physically meaningful amorphous structures.
    \textbf{b}, The denoiser model is trained to predict the displacement added to the amorphous structures. Displacements $\eps$ are sampled from Gaussian distributions, and added at train time to the training structures. Structure-level labels such as cooling rates are embedded to the training set with a Gaussian basis set.
    \textbf{c}, To generate new structures, the model is provided with a random input structure and a target cooling rate. The structure is denoised over multiple time steps following a noise schedule similar to the denoising diffusion probabilistic model (DDPM) framework.\cite{ho2020denoising}
    \textbf{d}, Figures of merit used to validate generated amorphous structures include short range order, medium range order, network connectivity, mechanical properties, and information entropy.
    \textbf{e}, In this work, the generative models for amorphous materials were used across multiple applications, including generating glassy structures conditioned to very slow cooling rate, large scales, pores, or reproducing the phase space of simulated and experimental metallic glasses.
    }
    \label{fig:overview}
\end{figure}

Here, we show that a graph neural network (GNN)-based diffusion model can faithfully generate amorphous structures across diverse compositions, densities, processing conditions, and data sources (Fig. \ref{fig:overview}).
Following a denoising diffusion probabilistic model (DDPM) framework\cite{ho2020denoising} for atomistic systems,\cite{Hsu2024Score} the GNNs are trained to denoise atomic environments added to known amorphous structures (Fig. \ref{fig:overview}a,b, \methods).
Later, trained models are used to sample new amorphous configurations from random distributions of atoms (Fig. \ref{fig:overview}a,c).
Generated amorphous structures recover short- and medium-range order, create distributions of outputs nearly indistinguishable from simulated structures, exhibit elastic and plastic behaviors compatible with their simulated counterparts, and can be conditioned on processing parameters such as cooling rates (Fig. \ref{fig:overview}d).
The methods are demonstrated with applications on amorphous silica (a-SiO$_2$), one simulated metallic glass, and one experimentally determined amorphous nanoparticle. The capabilities are used to generate large simulation cells, create processing-property relationships, produce porous structures, and augment experimental characterization data with synthetic results, enabling computational studies previously expensive or impractical (Fig. \ref{fig:overview}e).
These results provide a step toward the inverse design of amorphous materials and an efficient tool for sampling their vast configurational space.

\section{Results}

\subsection{Generating valid and novel amorphous materials}
\label{sec:unconditional}

\begin{figure}[htb]
    \centering
    \includegraphics[width = \textwidth]{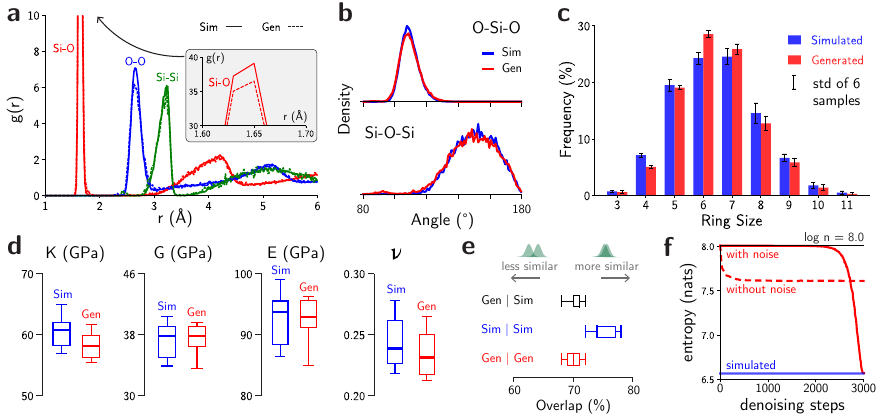}
    \caption{
    \textbf{Structures and properties of generated a-SiO$_2$ structures are nearly indistinguishable from simulated ones.}
    \textbf{a}, Comparison between the partial pair distribution functions (PDFs) of simulated (solid lines) and generated (dashed lines) shows that generated structures are quite similar to simulated ones.
    This close agreement is also seen in other structural metrics, including \textbf{b}, distributions of Si-O-Si and O-Si-O bond angles and \textbf{c}, ring size distributions.
    In \textbf{c}, the error bar is the standard deviation of across six simulated or generated samples.
    \textbf{d}, Elastic properties of generated structures are indistinguishable from the simulated ones. The properties computed here include the bulk modulus ($K$), shear modulus ($G$), Young's modulus ($E$), and Poisson's ratio ($\nu$).
    \textbf{e}, Generated structures follow the expected distribution of amorphous environments, as quantified by an information-theoretical strategy.\cite{schwalbekoda2025information}
    The overlap between generated (Gen) and simulated (Sim) distributions of environments (black) is nearly equal to the overlap between simulated and simulated (blue) distributions.
    In \textbf{d} and \textbf{e}, the  thicker line, box, and whiskers depict the median, interquartile range, and range of the distribution, respectively.
    \textbf{f}, Structures generated without a large external noise schedule cannot reproduce the information content of the reference structures, instead leading to structures with a high number of outlier environments (high entropy).
    On the other hand, a denoising process that includes an external noise schedule is able to generate more realistic structures.
    }
    \label{fig:validation}
\end{figure}

As a first case study, we trained a GNN-based diffusion model to generate a-SiO$_2$ structures sampled with MD simulations at a constant cooling rate of 1 K/ps (\methods).
Then, we verified whether generated structures displayed the correct short- and medium-range order for these materials.
Figure \ref{fig:validation} shows that generated structures exhibited excellent structural agreement with their simulated counterparts adopted as ground-truth.
Generated structures captured the short-range order characteristics of a-SiO$_2$, with partial PDFs and bond angle distributions (BADs) reproducing peak positions and intensities of the reference data (Fig. \ref{fig:validation}a).
At the medium range, Fig. \ref{fig:validation}c shows that generated structures reproduced ring distributions (\methods) with excellent accuracy, and only slightly overestimated the number of 6-membered rings.
Beyond structural distributions, which can average out outliers, macroscopic quantities such as mechanical properties are highly sensitive to unphysical environments, as they often lead to large variations in energy (see \supptext, Section \ref{sec:si:sensitivity}).
Thus, reproducing properties is a rigorous test to validate generative models, but have been largely underutilized within amorphous materials.
To enable this test, we calculated the stiffness tensor of simulated and generated structures according to the ground-truth potential used to produce the training data (see \methods\ for details).
Figure \ref{fig:validation}d shows that the elastic properties computed for the generated silica structures, including bulk modulus ($K$), shear modulus ($G$), Young's modulus ($E$), and Poisson's ratio ($\nu$), are statistically indistinguishable from their simulated counterparts.
In contrast with methods such as reverse Monte Carlo, where structural features are well-reproduced but properties can be severely wrong,\cite{mcgreevy2001reverse,maffettone2025can} our models produced structures in excellent agreement with reference simulation data across all relevant figures of merit.

As generative models learn a distribution of input data points, they are not only assessed according to the \textit{validity} of samples, but also according to their \textit{novelty}, thus verifying whether the model exhibits ``mode collapse.''\cite{goodfellow2014generative}
However, in contrast to crystalline structures or molecular graphs, novelty is challenging to define in the amorphous space (see Sec.\ \ref{sec:si:novelty} in the \supptext\ for a complete discussion).
We computed a probability distribution over environments for amorphous structures following an information-theoretical strategy,\cite{schwalbekoda2025information} and quantified the overlap between distributions.
This comparison between distributions resembles the Fr\'echet Inception Distance,\cite{heusel2017gans} which accounts for novelty in terms of distribution statistics rather than point-wise metrics.
Figure \ref{fig:validation}e shows that the distribution of overlap scores between different structure pairs is nearly identical.
Generated structures exhibited as much overlap with simulated structures as the overlap between two simulated structures, ruling out the existence of mode collapse across the system when a large external noise is used to generate structures.
On the other hand, Fig. \ref{fig:validation}f shows that the information entropy of generated samples is unreasonably large if no external noise is used during the denoising process (see \supptext, Sec. \ref{sec:si:noise}).
These results demonstrate that our diffusion model generates valid structures according to structural features\cite{comin2019deep,yong2024dismai,kwon2024Spectroscopy-guided} and property distributions, as well as novel structures, as quantified by their sampling distributions.

\subsection{Generating amorphous structures conditioned on processing parameters}

\begin{figure}[htb]
    \centering
    \includegraphics[width=\textwidth,height=0.8\textheight,keepaspectratio]{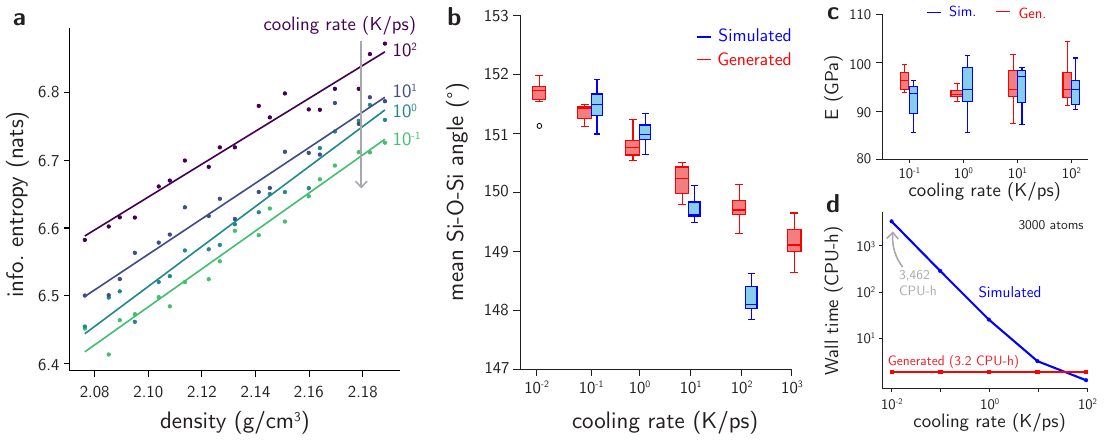}
    \caption{
    \textbf{Conditional generation of a-SiO$_2$ using diffusion models recover structural features and properties across cooling rates.}
    \textbf{a}, The diffusion models capture the correct trends in information entropy for amorphous structures generated under different cooling rates and densities. Faster rates often lead to structures with more outlier environments and thus higher information entropy.
    \textbf{b}, Trends in bond angle distributions are reasonably captured when a single diffusion model is used to generate structures across different cooling rates.
    The models continue to exhibit the correct trends even in regimes outside of the training domain (10$^{-2}$ and 10$^{3}$ K/ps)
    \textbf{c}, Young's modulus ($E$) of generated a-SiO$_2$ structures across different cooling rates.
    The moduli are highly sensitive to outliers, yet the denoiser is able to generate structures with accurate Young's modulus compared to simulations.
    In \textbf{b} and \textbf{c}, the horizontal line, box, and whiskers depict the median, interquartile range, and range of the distribution, respectively.
    The single data point in \textbf{b} ($10^{-2}$ K/ps) represents an outlier, which is away from the median by at least 1.5 times the interquartile range.
    \textbf{d}, The computational cost of simulating a-SiO$_2$ within the melt-quench process drastically increases with lower cooling rate, whereas the generative model has constant inference time.
    At very low cooling rates (10$^{-2}$ K/ps), the difference in computational wall time can reach 3 orders of magnitude.
    }
    \label{fig:conditional}
\end{figure}

Controlling processing conditions such as cooling rates in glassy materials is essential to model and design amorphous structures and their resulting properties.\cite{ vollmayr1996cooling, lane2015cooling,li2017cooling}
Given the cost of performing long MD simulations, cooling rates in computational models typically revolve around 10$^{12}$ K/s (or 1 K/ps), exceeding typical experimental cooling rates (1--100 K/s) by several orders of magnitude.
Thus, learning to sample structures at very low cooling rates can unlock the ability to produce increasingly realistic models for amorphous systems.
We trained our generative model on a-SiO$_2$ structures simulated with cooling rates between 10$^{-1}$ to 10$^{2}$ K/ps (\methods), then used an input to the GNN that conditioned the generation of amorphous structures to a target cooling rate (Fig. \ref{fig:overview}c).
Figure \ref{fig:conditional}a demonstrates that structures generated with higher cooling rate targets exhibited higher information entropy even at constant density, reproducing known relationships between configurational entropy and cooling rates in glasses (see Sec. \ref{sec:si:cooling-rate}).\cite{berthier2019configurational}
Moreover, Fig. \ref{fig:conditional}b shows that the dependence of average Si-O-Si angles with respect to the thermal history are reasonably captured in the generated a-SiO$_2$ structures, in agreement with simulated counterparts and the literature.\cite{vollmayr1996cooling, lane2015cooling,li2017cooling}
Beyond structural features, Fig. \ref{fig:conditional}c shows the Young's modulus ($E$) for simulated and generated a-SiO$_2$ structures across different cooling rates.
The values of Young's moduli remain consistent across cooling rates for both simulated and generated a-SiO$_2$ structures, ranging between 85--100 GPa.
This agreement is also true for all other elastic properties, including bulk modulus, shear modulus, and Poisson's ratio (Figs. \ref{fig:si:young_cooling}--\ref{fig:si:poisson_cooling}), confirming our ability to conditionally generate structures across cooling rates while reproducing their property distributions (see Sec. \ref{sec:si:cooling} in the \supptext\ for an extended analysis).

\begin{figure}[!h]
    \centering
    \includegraphics[width=1.0\linewidth]{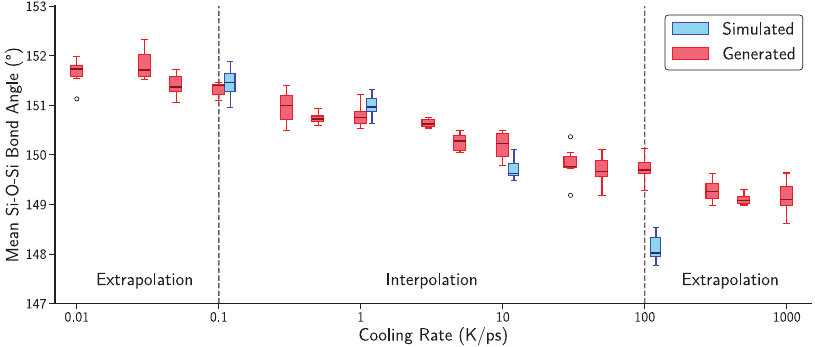}
    \caption{Mean Si–O–Si bond angle of a-SiO$_2$ structures as a function of cooling rate for simulated and generated samples.
    Each data point represents the average of six independent structures.
    The generated samples accurately reproduce the cooling rate dependence observed in simulated structures, with correct trends maintained even for cooling rates outside the training domain (10$^{-2}$ and 10$^3$ K/ps).
    }
    \label{fig:si:02-BAD-coolingrate-comparison}
\end{figure}

Remarkably, the trends in Figs. \ref{fig:conditional}b,c generalize to both faster (10$^{3}$ K/ps) and slower (10$^{-2}$ K/ps) cooling rates beyond the training regime (see Fig. \ref{fig:si:02-BAD-coolingrate-comparison} for more examples), which can be used to generate structures at low cooling rates that are expensive for traditional simulations.
To illustrate the impact of this observation, Fig. \ref{fig:conditional}d compares the estimated computational cost (in CPU-h) for simulating amorphous structures with 3000 atoms across cooling rates compared to the time required to evaluate the generative model (see \methods\ for hardware details).
As slow cooling rates can only be simulated by longer MD simulations, the computational cost of simulating realistic glasses follows a power law with the cooling rate, while the generative model produces structures at constant time.
At slow cooling rates such as 0.01 K/ps, the diffusion model produces a-SiO$_2$ structures at least 1000 times faster (in CPU-h) than MD simulations (Fig. \ref{fig:conditional}d) for a system with 3000 atoms.
Performing melt-quench simulations for the largest system in this work (112,848 atoms, Sec. \ref{sec:fracture}) at 0.01 K/ps would require about 89,702 CPU-h using MD in our hardware (\methods), a substantial investment of computational resources.
In practice, we use GPUs to generate structures with the GNN, further reducing the wall times.
Even when compared to GPU-accelerated MD, the generative process drastically improves the effectiveness of sampling amorphous structures at low cooling rates and large sizes, demonstrating that the correct processing-structure-property relationships can be captured with the model.

\subsection{Fracture tests with large and slow-cooled amorphous silica}
\label{sec:fracture}

\begin{figure}[htb]
    \centering
    \includegraphics[width=\textwidth,height=0.6\textheight,keepaspectratio]{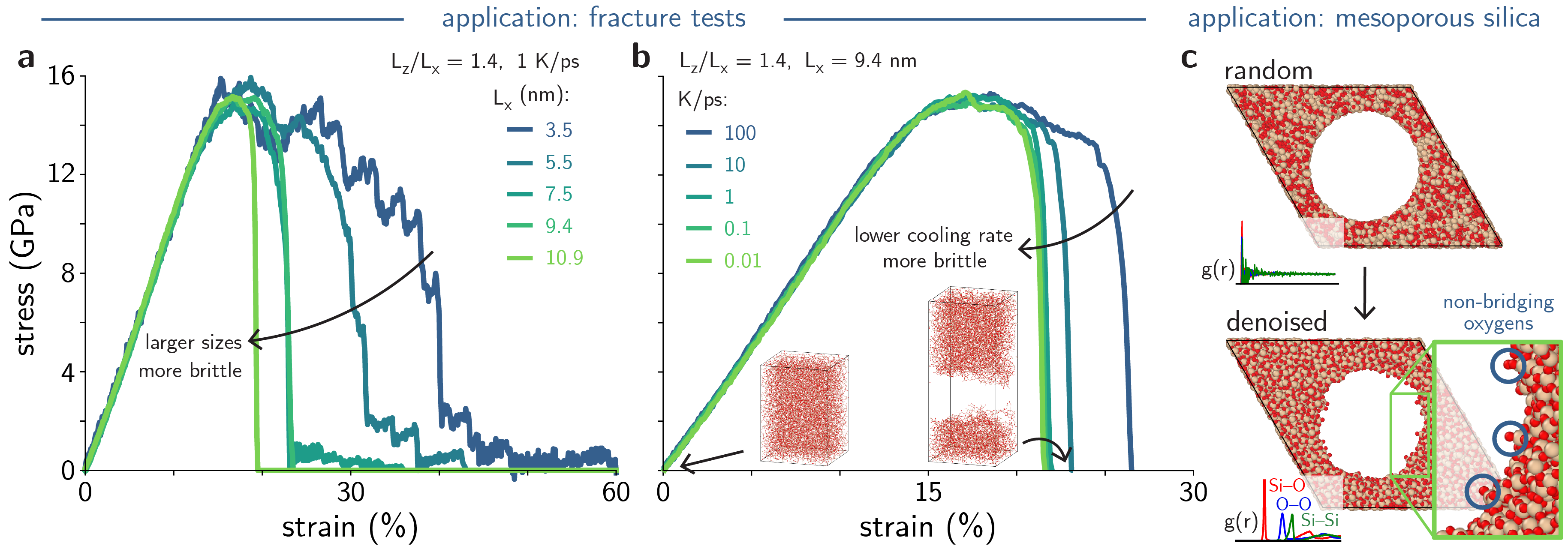}
    \caption{\textbf{Shape- and size-agnostic amorphous structure generation enables simulations on fracture and structural porosity.}
    \textbf{a}, Stress-strain curves show that generated a-SiO$_2$ structures recover the right behavior of elastic and plastic deformation behavior across length scales, with more ductile fracture at smaller length scales while maintaining a constant ultimate tensile strength at fixed aspect ratio ($L_z / L_x = 1.4$).
    \textbf{b}, Stress-strain curves from generated a-SiO$_2$ structures of the same shape and size ($L_z / L_x = 1.4$, $L_x = 9.4$ nm) under conditional cooling rates show that slower cooling rate leads to a more brittle behavior for amorphous silica.
    \textbf{c}, The model successfully generates mesoporous a-SiO$_2$ structures despite not being trained on surfaces or porous structures.
    Partial pair distribution functions (PDFs) indicate that the structure remains amorphous despite the presence of surfaces and pores.
    Interestingly, the pore surface reproduces the experimental concentrations of non-bridging oxygens, holding promise to modeling silanol defects and other interfacial behavior of amorphous mesoporous silica.
    }
    \label{fig:fracture}
\end{figure}

\begin{figure}[!h]
    \centering
    \includegraphics[width=1.0\linewidth]{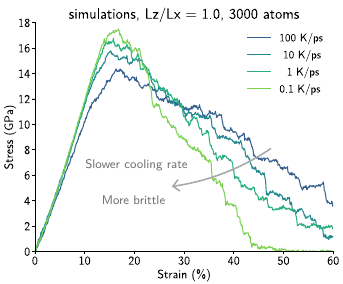}
    \caption{Stress-strain curves from simulated a-SiO$_2$ structures under different cooling rates show that slower cooling rate leads to a more brittle behavior.
    The initial structure was a cube with a total of 3000 atoms, and the fracture simulation was performed as described in the Methods.
    These results show that the ductility of the glassy system is high at this small scales, in contrast with results at large scales.
    }
    \label{fig:si:fracture_simu_coolingrate}
\end{figure}

As our generative model samples structures for any input size or shape and for a range of cooling rates, we investigated the fracture behavior of generated a-SiO$_2$ structures for different length scales and processing conditions.
Experimental observations have shown that a-SiO$_2$ exhibited a brittle-to-ductile transition at the nanoscopic regime,\cite{celarie2003glass, luo2016size} while computational studies were typically limited by small sizes and fast cooling rates.\cite{vollmayr1996cooling,lane2015cooling,yuan2012molecular}
We used our diffusion model to generate a-SiO$_2$ with different sizes and cooling rates, and performed fracture test simulations for each of them (\methods).
Figure \ref{fig:fracture}a shows the stress-strain curves obtained for a-SiO$_2$ structures generated with a fixed aspect ratio of $L_z$/$L_x$ = 1.4, as $z$ being the direction of the tensile strain, and $L_x$ varying from 3.5 nm (3,735 atoms) to 10.9 nm (112,848 atoms).
Generated structures reproduced the increasingly brittle behavior of amorphous silica at large length scales, while smaller simulation lengths ($L_x < 6$ nm) were more ductile (\supptext, Sec. \ref{sec:si:fracture}).
The influence of low cooling rate effects on fracture, however, is often not studied at large length scales due to its associated computational cost.\cite{lane2015cooling,zhang2020critical}
On the other hand, our generative model allows sampling these large, slow-cooled structures at minimal computational cost.
Figure \ref{fig:fracture}b compares the behavior of slow- and fast-cooled structures for a-SiO$_2$ structures generated at large sizes ($L_z$/$L_x$ = 1.4, $L_x$ = 9.4 nm) and different target cooling rates.
The results show that lower cooling rates decreased the ductility of the amorphous material,\cite{fan2017effects} but up to a certain limit.
The fracture strain for samples generated with cooling rates of 0.1 and 0.01 K/ps are nearly identical even though the structures are slightly different (Fig. \ref{fig:conditional}b), showing how the fracture behavior converges at very low cooling rates.
This contrasts with the continuous trend observed in simulated glasses at much smaller system sizes of 3000 atoms, for which the ductility was drastically overestimated and continued to decrease at low cooling rates (Figs. \ref{fig:si:fracture_simu_coolingrate},\ref{fig:si:fracture_cube}).

\subsection{Generating mesoporous amorphous structures}

Mesoporous silica particles exhibiting pores with diameters between 2--50 nanometers are ubiquitous in many applications ranging from catalysis and separation to drug delivery. \cite{kresge1992ordered,tang2012mesoporous}
Despite their wide uses in experiments, computational sampling of these structures is laborious and can lead to incorrect pore structures or defect concentrations.\cite{calderon2018structure, fought2022modeling}
We used our generative approach to produce a porous a-SiO$_2$ structure by denoising a structure from atoms sampled in a hexagonal lattice with a cylindrical pore (\methods, Fig. \ref{fig:fracture}c).
Despite never been trained on surfaces or porous structures, our diffusion model generated excellent amorphous surfaces and a wall that exhibited the right short-range ordering.
Interestingly, generated porous structures correctly captured the atomic structure at the pore surface, comprised mostly of pristine SiO$_4$ tetrahedra and non-bridging oxygens (NBO).
The concentrations of NBOs were also in agreement with typical ranges of 2-4 NBOs/nm$^2$ from experimental data,\cite{zhao1997comprehensive, kozlova2010post} and were a function of the initial wall density used as starting configuration for the denoiser.
At higher densities, the pore tended to shrink and form more densed NBOs at the pore wall, while low-density walls tended to form less NBOs at the surface (Fig. \ref{fig:si:pore-sio2}).
As NBOs are responsible for a variety of properties of mesoporous silica in the presence of hydrogen, such as catalytic or hydrophobicity, obtaining these substructures directly from the generative approach can drastically accelerate computational modeling in these fields.

\subsection{Generating metallic glasses}

\begin{figure}[ht!]
    \centering
    \includegraphics[width=\linewidth]{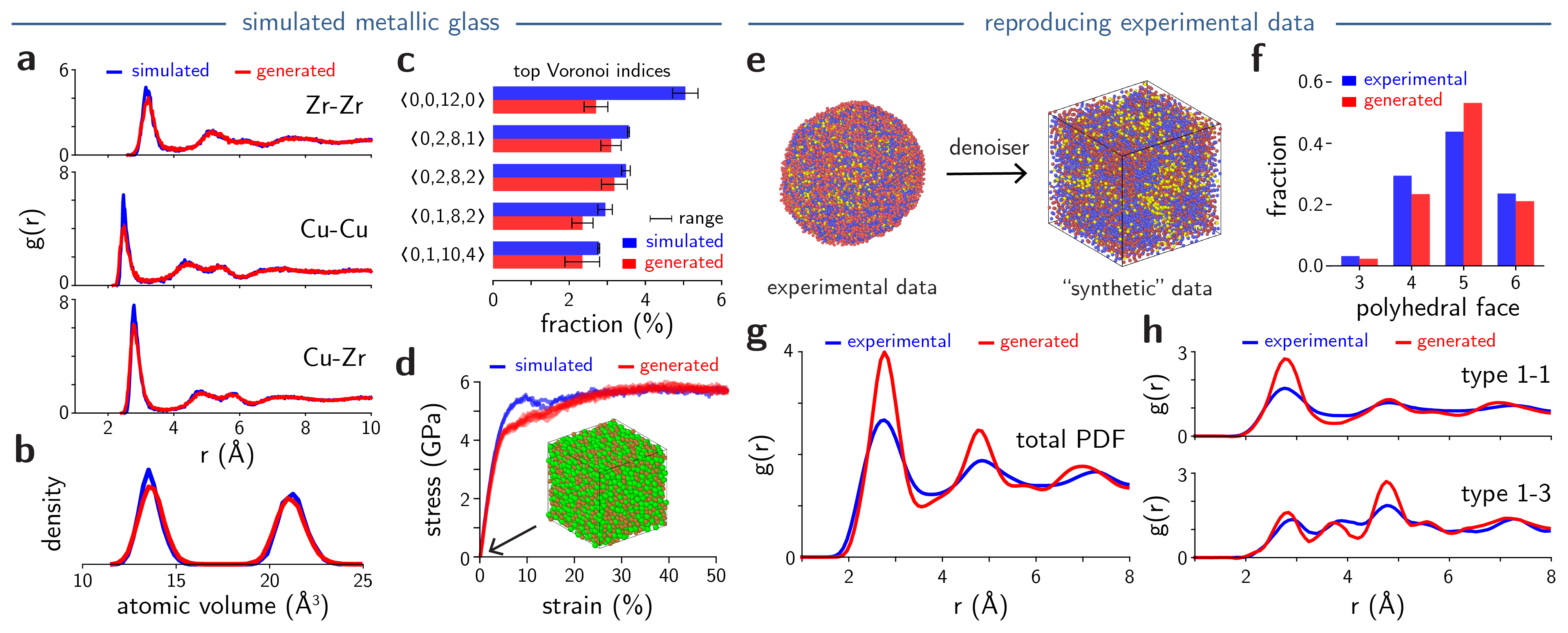}
    \caption{\textbf{Generating metallic glass structures from simulated and experimental data sources.}
    \textbf{a}, Partial pair distribution functions (PDFs) $g(r)$ of generated Cu$_{50}$Zr$_{50}$ structures (red lines) exhibit good agreement with PDFs of their simulated counterparts (blue lines), including capturing the shorter Cu-Cu pair interactions compared to Cu-Zr and Zr-Zr pairs.
    \textbf{b}, Distributions of atomic volume exhibit great agreement between the coordinations and local environments of generated Cu$_{50}$Zr$_{50}$ structures (red) and their simulated, ground-truth counterparts (blue).
    \textbf{c}, Generated Cu$_{50}$Zr$_{50}$ structures  reproduce the fractions of major Voronoi polyhedra compared to their simulated counterparts, only under-representing $\langle 0,0,12,0 \rangle$.
    Error bars for indices indicate the range of the distribution.
    \textbf{d}, Stress-strain behavior of generated Cu$_{50}$Zr$_{50}$ structures (red, 5 samples) approach the ones from their simulated Cu$_{50}$Zr$_{50}$ counterparts (blue, 2 samples), with equivalent Young's modulus and strength, despite slightly underestimated yield strength.
    \textbf{e}, Beyond simulation data, the generative diffusion model can be trained on experimental data, such as the atomistic structure characterized by Yang \textit{et al.},\cite{yang2021Determininga} and generate synthetic counterparts to the experimental results.
    \textbf{f}, Generated atomic structure using experimental data shows good agreement on polyhedral face fractions (red) compared to results from the experimental atomic model (blue).
    \textbf{g}, Total PDFs and \textbf{h}, partial PDFs show that generated structures correctly capture the pair interactions from the experimental three-component metallic glass, including the anomalous peak positions and intensities for the 3-3 pairs.
    }
    \label{fig:metals}
\end{figure}

To extend our generative approach beyond a-SiO$_2$, we tested our methodology against Cu-Zr metallic glass structures and properties, a non-covalent amorphous system with a challenging potential energy surface.
Amorphous Cu-Zr structures with a fixed composition of Cu$_{50}$Zr$_{50}$ and 5000 atoms were obtained from Ref. \citenum{wang2020cuzr} to train a diffusion model without conditional targets.
Then, given a box with randomly sampled atoms at a chosen composition and density, we used the diffusion model to generate the structure of the metallic glass, which can be followed by a short, post-denoising MD equilibration (\methods).
Figure \ref{fig:metals}a shows that the partial PDFs between simulated and generated Cu$_{50}$Zr$_{50}$ glasses are in excellent agreement with each other, with generated structures showing near-perfect peak positions and only slightly lower intensities compared to the reference structures.
This is true even for long-range peaks around 7 and 9 \AA, which were outside of the cutoff of the GNN (5 \AA), but were treated indirectly by the message-passing layers.
The atomic volume in Fig. \ref{fig:metals}b and fractions of Voronoi indices in Fig. \ref{fig:metals}c for simulated and generated Cu$_{50}$Zr$_{50}$ also show good agreement with each other and only underestimated $\langle 0,0,12,0 \rangle$ icosahedral motifs in generated samples.
Given the sensitivity of Voronoi polyhedra to the structural fidelity, the ability of the model to correctly recover these trends is remarkable.
Furthermore, Fig. \ref{fig:metals}d shows that the elastic and plastic regimes of the generated amorphous Cu-Zr structures (6 independent samples) are also compatible with the reference curves (2 samples).
Though the yield stress is slightly underestimated in the generated structures, mirroring the results in Fig. \ref{fig:si:fracture_cube}, the strength curves of both metallic glasses are within good agreement with each other, especially when considering the sensitivity of stress-strain curves from these metallic glasses depending on the sample preparation\cite{nawano2023characterizing} and the limited training data used for the denoiser model.
This proves that the generative model is not limited to amorphous silica networks, but can also produce valid amorphous structures and properties for metallic glassy systems.

\subsection{Augmenting experimental characterization data on amorphous structures}
\label{sec:experimental}

\begin{figure}[!h]
    \centering
    \includegraphics[width=1.0\linewidth]{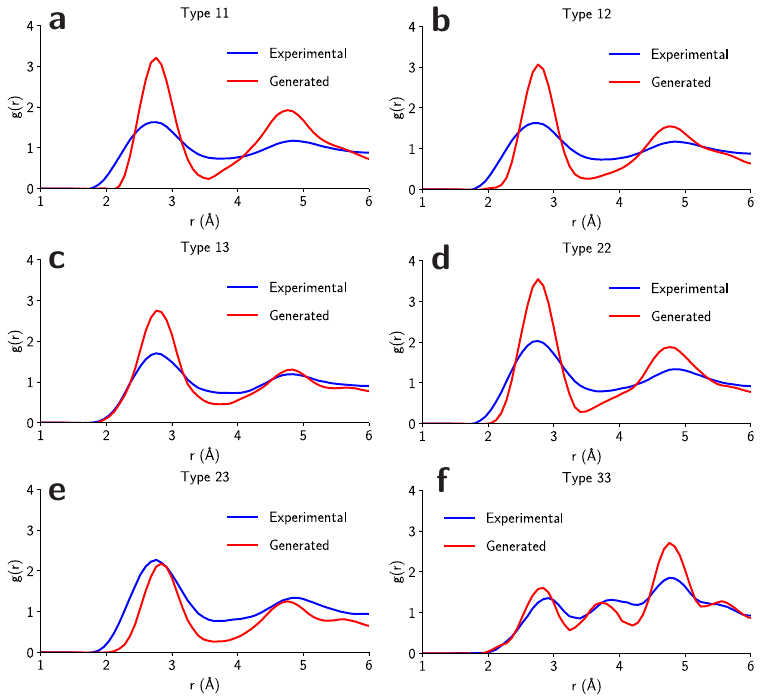}
    \caption{Comparison of partial pair distribution functions $g(r)$ between the experimental model and denoiser generated model of amorphous nanoparticle.
    The experimental nanoparticle contains 3 types of atoms, and denoiser model can well reproduce the short-range order of type \textbf{a}, 1-1, \textbf{b}, 1-2, \textbf{c}, 1-3, \textbf{d}, 2-2, \textbf{e}, 2-3, and \textbf{f}, 3-3 pairs.
    }
    \label{fig:si:partialPDF_nano}
\end{figure}

Given that characterization data on amorphous structures is challenging to obtain at high resolutions, generative models can be used to augment experimental datasets with realistic, synthetically generated data.
We trained our model on the atomistic structure of a multi-component metallic glass from Yang \textit{et al.}\cite{yang2021Determininga} obtained using atomic electron tomography (AET) (Fig. \ref{fig:metals}e).
The data contained a finite metallic glass nanoparticle with three different element classes denoted as types 1, 2, and 3.
Generated structures reproduced the distribution of polyhedral faces from the experimental data (Fig. \ref{fig:metals}f, see also \supptext, Sec. \ref{sec:si:voronoi}), indicating that the model captured the correct geometrical distribution of the data.
Furthermore, Figs. \ref{fig:metals}g,h show that the diffusion model approximated the total PDF of the experimental data at realistic levels despite being trained on a single example of experimental structure.
Discrepancies between the two PDFs in Fig. \ref{fig:metals}g, especially at longer ranges, are due to the mismatch between the radial data computed for a finite particle in the experimental data and the generated bulk structure represented with periodic boundary conditions.
Nevertheless, the generated structure reproduced the experimental behavior observed in the medium-range, where certain substructures were shown to form crystal-like ``superclusters'' that created medium-range order for specific atom pairs, and atypical PDFs for specific atom pairs.\cite{yang2021Determininga}
Figure \ref{fig:metals}h illustrates that the model correctly captured the correct short-range ordering and pairwise distributions in the experimental data (see Fig. \ref{fig:si:partialPDF_nano} for all partial PDFs).
In typical pair distribution functions such as that for 1-3 pairs in Fig. \ref{fig:metals}h, generated structures exhibited the right peak positions and their relative intensities of the PDF, including medium-range features between 5--6 \AA.
Remarkably, even for atypical partial PDFs such as the ones for 3-3 pairs in the experimental data, the model captured the correct distribution of atomistic species, as demonstrated by the agreement in peak positions around 2.9, 3.8, and 4.8 \AA, with the third peak being the most prominent one.
At medium-range, the model also captured the shoulder in the PDF between 5.5--6.5 \AA, and the longer-range peak around 7.3 \AA.
This shows that the generative model can quantitatively reproduce experimental datasets and can be used to produce synthetically generated structures that mimic the experimental results at arbitrary sizes and shapes, or assist in the development of atomistic models for structural characterization.

\section{Discussion}

Our generative diffusion model substantially advances computational modeling of amorphous materials, accurately producing physically realistic structures and properties across various compositions, data sources, and processing conditions.
Unlike previous generative methods for amorphous materials, whose unphysical artifacts or constrained structural diversity limited their broad use in simulations, our model consistently replicates key short- and medium-range structural characteristics and mechanical properties, highlighting its robustness and reliability.
Importantly, this framework offers efficient computational performance, drastically reducing the computational cost typically associated with molecular dynamics simulations.
Our work provides a roadmap not only for the training process of diffusion models for amorphous materials, but also how to quantitatively assess model quality in terms of structural novelty, properties, and applications.

As with any model, our approach exhibits limitations. Occasional outlier environments, although rare, require short MD simulations to refine the generated structure.
Furthermore, though several properties are predicted with high accuracy, yield strengths are still slightly underestimated, showing this metric may be a good indicator of model quality within generative models for amorphous structures as the field progresses.
Future research can extend this approach toward broader classes of complex materials, including metallic or semiconducting alloys or polymers.
Additionally, integrating generative methods with experimental characterization can automate structural elucidation, particularly in fields where experimental data is scarce or costly to obtain.
The adaptability and diversity of applications demonstrated here suggest that generative diffusion models can already be used in a range of computational materials science problems, accelerating their modeling and discovery.

\section*{Methods}
\customlabel{sec:methods}{Methods}

\subsection*{Generative diffusion model} \label{diffusion model}

\noindent\textbf{Intuition of the generative diffusion model for amorphous structures:}
To create a generative model for amorphous structures, we adapted the denoising diffusion probabilistic models (DDPM) approach to amorphous materials by learning a mapping between a random distribution of atoms and a physically meaningful amorphous structure using a graph neural network (GNN).\cite{ho2020denoising, Hsu2024Score}
Figure \ref{fig:overview}a exemplifies the training and generative denoising processes for amorphous SiO$_2$ (a-SiO$_2$), with their partial pair distribution functions (PDFs) illustrating how short- and medium-range order varies across the system.
The forward diffusion process (left to right in Fig. \ref{fig:overview}a) gradually introduces Gaussian noise to atomic positions from simulated amorphous structures until the final structure resembles a random distribution of particles.
After training, the model is used to generate amorphous structures from randomly sampled particles with the denoising process (Fig. \ref{fig:overview}a, right to left).
Given a target cooling rate and an initial configuration with fixed density and shape, the model sequentially denoises the configuration in a process that resembles ``Langevin dynamics'' with a learned score.
This process adds an extra noise at each denoising step, which acts as an effective ``temperature'' for the system and escaping local minima in the learned score function (see Sec. \ref{sec:si:noise} in the \supptext).
Within the DDPM, this extra noise follows a schedule and becomes progressively smaller throughout the denoising process.\cite{ho2020denoising}

\noindent\textbf{Score-based denoiser for amorphous structures:}
Mathematically, a generative model approximates the distribution $p(\x)$ over atomic structures $\x$ of the material system of interest.
Under this DDPM strategy for atomistic structures, we train a model to approximate the distribution of atomic environments of amorphous configurations $p(\x)$ inferred from a dataset of known configurations.
This approximation of $p(\x)$ is first performed by considering an input structure $\x_0$ and defining a Markov chain process such that

\begin{equation}
    \x_0 \rightarrow \x_1 \rightarrow \ldots \rightarrow \x_T,
\end{equation}

\noindent where $T$ is the number of diffusion steps.
This forward process is defined by a distribution $q(\x_t | \x_{t-1})$, $1 \leq t \leq T$.
For simplicity, the distribution $q$ is often chosen as a Gaussian distribution, where

\begin{equation}
    q(\x_t | \x_{t-1}) = \mathcal{N} \left(
        \boldsymbol{\mu}_t (\x_{t-1}), \sigma_t^2 \mathbb{I}
    \right),
\end{equation}

\noindent with $(\boldsymbol{\mu}_t, \sigma_t)$ functions of the denoising timestep $t$ and $\mathbb{I}$ the identity matrix.
Within the diffusion model framework, given that the ground-truth distribution $p$ is often intractable,  a model $p_\theta$ with parameters $\theta$ is trained to approximate, at each step $t$, the score function of the noise distribution $q(\x_t | \x_{t-1})$,\cite{hyvarinen2005estimation}

\begin{equation}\label{eq:loss}
\min_\theta \mathbb{E}_{t} \left[
    \sigma^2_t \mathbb{E}_{\x_0 \sim \mathcal{D}} \mathbb{E}_{q(\x_t | \x_0)} \left[
        \lVert
        p_\theta (\x_{t}, t) - \nabla_{\x_t} \log q (\x_t | \x_0)
        \rVert^2
    \right]
\right].
\end{equation}

\noindent Within this notation, $p_\theta (\x_{t}, t)$ is called the score model, and the expectation $\mathbb{E}_{\x_0 \sim \mathcal{D}}$ is performed over the dataset $\mathcal{D}$ of initial (training) structures $\x_0$.

When the distribution $q$ is chosen to be Gaussian with $\boldsymbol{\mu}_t = 0$ and we define $\x_t = \x_{t-1} + \sigma_t\boldsymbol{\eps}$, $\boldsymbol{\eps} \sim \mathcal{N}(\mathbf{0}, \mathbb{I})$, its score can be written directly as

\begin{equation}\label{eq:score}
\nabla_{\x_t} \log q (\x_t | \x_{t-1}) = -\frac{\boldsymbol{\eps}}{\sigma_t},
\end{equation}

\noindent which defines a per-atom displacement $\boldsymbol{\varepsilon}_t$ that maps the noisy atomic positions $\x_t$ to its previous, denoised state $\x_{t-1}$.\cite{ho2020denoising}
By redefining the model $p_\theta (\x_{t}, t)$ as $\boldsymbol{\eps}_\theta (\x_{t}, t) = -\sigma_t p_\theta (\x_{t}, t)$ and adopting a denoising score-matching loss,\cite{vincent2011connection} which simplifies the objective function of Eq. \eqref{eq:loss} to

\begin{equation}\label{eq:loss-2}
\min_\theta \mathbb{E}_{t} \left[
    \mathbb{E}_{\x_0 \sim \mathcal{D}} \mathbb{E}_{q(\x_t | \x_{t-1})} \left[
        \left\lVert
        \boldsymbol{\eps}_\theta (\x_{t}, t) - \boldsymbol{\eps}
        \right\rVert^2
    \right]
\right].
\end{equation}

\noindent Equation \eqref{eq:loss-2} indicates that a model $\boldsymbol{\eps}_\theta$ is given an input structure $\x_t$ and is trained to predict the noise $\boldsymbol{\eps}$ that maps from the denoised state $\x_{t-1}$ into $\x_t$ given a noise schedule $\sigma_t$.

\noindent\textbf{Model architecture:}
Following previous approaches,\cite{Hsu2024Score} the denoiser model uses a customized version of the E(3)-equivariant NequIP model\cite{batzner2022nequip} to implement the training from in Eq. \eqref{eq:loss-2}.
The model utilizes an initial embedding layer that maps atomic species information onto an 8-dimensional scalar feature space (\texttt{8x0e}) for both node attributes and atomic numbers, with interactions truncated at the specified cutoff radius (5 \AA).
The network architecture consists of three convolution layers (\texttt{num\_convs = 3}) within the message-passing framework.
Hidden layer representations are configured as 64 scalar features and 32 vector (rotation-equivariant) features.
Edge features are represented with \texttt{4x0e + 4x1o + 2x2e} features encoding scalar, vector, and rank-2 tensor components.
The radial basis functions are parameterized through a two-layer neural network with 16 and 64 neurons, respectively, to model distance-dependent interactions.
The model considers up to 12 nearest neighbors per atom (\texttt{num\_neighbors = 12}) and produces vector outputs, ensuring that the predicted noises are equivariant under rotations and translations while maintaining the fundamental symmetries of the physical system.

\noindent\textbf{Cooling rate embedding:} \label{cooling rate embedding}
To incorporate cooling rate information into the model architecture, we implemented a Gaussian basis embedding module, which maps a scalar cooling rate value into a high-dimensional feature vector.
Specifically, we project each cooling rate onto a set of 9 Gaussian basis functions with means distributed uniformly across the normalized cooling rate range in a $\log_{10}$ scale, thus mapping the interval of cooling rates from $[10^{-2}, 10^{3}]$ K/ps to $[-2, 3]$.
Each basis function's activation represents the similarity between the input cooling rate and that basis function's mean value, effectively creating a smoothly varying representation.
These activations are then processed through a two-layer neural network with a SoftPlus nonlinearity, producing a dense embedding vector that captures the cooling rate's influence on structural properties.
This embedding is summed to the hidden representation at each convolution layer of the GNN.

\noindent\textbf{Training procedure:}
the model was trained on a single NVIDIA RTX A6000 GPU.
The per-coordinate, per-atom noise perturbation $\eps$ was added at train time and sampled from a normal distribution $\mathcal{N}(0,\sigma)$, where $\sigma$ ranged from 0.001 to 0.75 \AA{} based on the noise schedule.
At train time, the schedule $t$ is sampled from a uniform distribution, where $1 \leq t \leq T$.
Maximum displacement parameters were calibrated to approximately 50\% of typical bond lengths, ensuring that the diffusion process leads to fully random or highly noised configurations.
Cooling rates associated with individual training configurations were embedded using a Gaussian basis set, as details described in the above section.
The conditional denoiser model was trained for 40,000 iterations with a batch size of 16 and learning rate of $2\times 10^{-4}$, with validation loss monitored to assess convergence.
For unconditional denoiser models (CuZr and experimental nanoparticle systems), the number of training iterations was 20,000.

\noindent\textbf{Generative denoising process:} \label{method:inference}
The number of atoms and lattice parameters were selected as inputs based on desired density, lattice shape, and atom types.
All atoms were initially sampled from a uniform distribution within the simulation cell, meaning each of their fractional coordinates $(f_x, f_y, f_z)$ was sampled from $\mathcal{U}(0, 1)$.
The denoiser model was applied to generate atomic structures from these initialized configurations with extra noise (Fig. \ref{fig:overview}c).
The per-atom, per-coordinate extra noise is sampled from a normal distribution $\mathcal{N}(0,\sigma_t)$ with mean 0 and standard deviation $\sigma_t$ represents the noise magnitude following a noise schedule dependent on the denoising step $t$.
The noise magnitude was decreased linearly to zero over the extent of 2,900 generation steps.
Thus, each initial configuration underwent 2,900 denoising steps with extra noise (with $\sigma_{t=0}$ = 1.0) followed by 100 denoising steps without extra noise.
Despite being denoised from the same initial structure, adding ``large'' initial extra noise and using different random seeds ensures that generated structures are different from each other.

\noindent\textbf{Post-generation MD refinement:}
Generated configurations were subsequently refined using molecular dynamics simulations consisting of 25 ps in the \textit{NVT} ensemble followed by 25 ps in the \textit{NPT} ensemble using the potential and settings for MD described below.
This last-step MD refinement was performed for a-SiO$_2$ and Cu$_{50}$Zr$_{50}$ systems but was omitted for structures used to reproduce experimental data.

\noindent\textbf{Density dependence on the cooling rate:}
For conditional denoising based on different cooling rates, simulation boxes are kept constant during the denoising process, but the cooling rate is known to influence the density.
To account for this fact, we extracted densities from simulated structures across different cooling rates and identified a reasonably linear correlation between a-SiO$_2$ density and the logarithm of the cooling rate (see Fig. \ref{fig:si:02-density-coolingrate-comparison} and \supptext, Sec. \ref{sec:si:cooling}) within the ranges of interest.\cite{vollmayr1996cooling,lane2015cooling}
Thus, when selecting the targeted cooling rate, we fix the density of the initial, randomly sampled configuration using a linear regression model valid within the range of cooling rates of interest.
Future work can focus on sampling amorphous structures with variable densities throughout the denoising process.

\noindent\textbf{Denoising a porous structure:}
To create an amorphous a-SiO$_2$ structure, we performed the denoising process described above with a different initial condition, where atoms were sampled according to a non-uniform distribution throughout the lattice.
Atoms were sampled from a random distribution over a hexagonal lattice with predefined wall density, number of atoms, and lattice length.
However, an excluded volume in the shape of a cylinder prevented atoms from being sampled within the ``pore'', thus mimicking a uniform atomic distribution of atoms for all regions outside the pore (Fig. \ref{fig:fracture}c).
Then, the porous structure was denoised following the exact same procedure described above, with a target cooling rate of 1 K/ps.

\subsection*{Molecular dynamics simulations} \label{md simulations}

\noindent All molecular dynamics (MD) simulations were performed using the Large-scale Atomic/Molecular Massively Parallel Simulator (LAMMPS) software\cite{Thompson2022LAMMPS} (v. 17 Apr 2024).
A fixed integration timestep of 1.0 fs was used throughout all simulations.

\noindent\textbf{SiO$_2$ simulations:}
To generate training data for the SiO$_2$ systems, we conducted MD simulations of amorphous silica comprised of 3000 atoms across multiple cooling rates.
We executed six independent simulations at each cooling rate, maintaining the same settings while varying initial configurations.
We adopt the Jakse interatomic potential,\cite{bouhadja2013structural} parametrized from \textit{ab initio} calculations, to model a-SiO$_2$.
Short-range interactions are calculated with an 8.0 \r{A} cutoff.
Coulomb interactions are calculated by adopting the Fennell damped shifted force model with a damping parameter of 0.25 \r{A}$^{-1}$ and a global cutoff of 8.0 \r{A}.
Periodic boundary conditions were applied in all three directions during MD simulations.

All a-SiO$_2$ samples were prepared via conventional melt-quench method.
Initial structures were generated by randomly placing SiO$_2$ molecules in cubic boxes using PACKMOL (v. 20.13.0),\cite{martinez2009packmol} enforcing a 2.0 \r{A} minimum separation between molecules to prevent unphysical overlaps.
Following energy minimization, configurations underwent sequential 50 ps relaxations in the canonical (\textit{NVT}) and isothermal-isobaric (\textit{NPT}) ensembles at 300 K.
Complete melting was achieved with a 50 ps-long \textit{NPT} equilibration at 5000 K and zero pressure, eliminating the ``memory'' of the initial configuration.
Subsequently, liquid systems were cooled from 5000 to 300 K under zero pressure \textit{NPT} conditions at rates of 10$^{-1}$, 10$^0$, 10$^1$, and 10$^2$ K/ps.
The obtained glass structures are further relaxed one last time at 300 K for 100 ps in the \textit{NPT} ensemble.
This process follows the standards from previous work\cite{li2017cooling} shown to reproduce multiple experimental features from a-SiO$_2$ glasses.

\noindent\textbf{CuZr simulations:}
We adopted the published dataset from Wang \textit{et al.}\cite{wang2020cuzr}, which used a set of optimized embedded-atom method (EAM) potentials to simulate CuZr metallic glass models.\cite{cheng2009cuzrEAM}
Specifically, we used two of their Cu$_{50}$Zr$_{50}$ configurations, each containing 5,000 atoms, to train our denoiser model.
The potential from Cheng \textit{et al.}\cite{cheng2009cuzrEAM} was later used to perform the short MD refinement and the fracture test (described below) for generated CuZr structures.
The post-denoising MD simulation was 25 ps-long at the \textit{NVT} ensemble, followed by a 25 ps-long simulation at the \textit{NPT} ensemble, both at 300 K.

\noindent\textbf{Computation of elastic properties:}
the stiffness tensor $C_{ij}$ of the equilibrated glasses is computed by performing a series of 6 deformations (i.e., 3 axial and 3 shear deformations along the 3 axes) and calculating the curvature of the potential energy $U$:

\begin{equation}
C_{ij} = \frac{1}{V} \frac{\partial^2 U}{\partial e_i\partial e_j},
\end{equation}

\noindent where $V$ is the glass volume,  $e$ is the strain, and $i,j$ are the indices corresponding to each Cartesian direction.
All simulated configurations exhibited near-complete isotropy.
Bulk modulus (K), shear modulus (G), Young's modulus (E), and Poisson's ratio ($\nu$) are then calculated from the stiffness tensors.

\noindent\textbf{Fracture tests:}
fracture simulations are performed on different starting bulk configurations by deforming the structures along a single direction during an MD simulation.
Uniaxial deformation is achieved by imposing a constant strain rate of 10$^9$ s$^{-1}$ along the $z$ direction while allowing lateral dimensions ($x$ and $y$) to relax freely under zero lateral pressure, approximating realistic uniaxial tension conditions.
The deformation process is controlled using an \textit{NPT} ensemble maintained at 300 K and 1 bar.
True stress is calculated from the negative $z$-component of the stress tensor.

\subsection*{Computational cost estimates} \label{method:computation}

We used a single CPU core of Intel(R) Xeon(R) CPU E5-2650 v4 @ 2.20GHz from UCLA's Hoffman2 Cluster to estimate the computational cost of MD simulation and denoiser generation for 3000 atoms a-SiO$_2$.
The entire MD simulation process includes initialization, melting, cooling and relaxation parts as detailed above, where only the cooling part is affected by the choice of cooling rate.
The denoising process (inference) consists of a total of 3000 denoising steps (2900 steps with extra noise and 100 steps without extra noise).
The trained denoiser model is also saved, loaded, and evaluated on the same single CPU for this comparison.
In practice, we used GPUs for the generative inference process, which perform faster than using CPUs for inference of the GNNs.

To estimate the computational cost of traditional melt-quench MD simulations for large systems at slow cooling rates, we benchmarked structures at target sizes for 25 ps.
All simulations include fixed computational overhead include 50 ps \textit{NVT} initialization, 50 ps \textit{NPT} initialization, 50 ps \textit{NPT} melting, and 100 ps \textit{NPT} relaxation (250 ps total).
The quenching process duration varies with cooling rate, requiring 47, 470, 4,700, 47,000 and 470,000 ps for 10$^2$, 10$^1$, 10$^0$, 10$^{-1}$, and 10$^{-2}$ K/ps, respectively.
Total computational cost is calculated by multiplying the average time per step obtained for a 25 ps MD simulation by the total number of steps to perform the simulation.
For the large system estimate, the cost of performing a 25 ps MD simulation of 112,848 atoms within a-SiO$_2$ and the simulation settings described above is 4.77 CPU-h.
This results on the total estimated cost of 89,702 CPU-h for the same size under 0.01 K/ps cooling rate described in the main text.

\subsection*{Structural analysis}

All structural features of simulated a-SiO$_2$ structures were performed with six independent samples of 3000 atoms each.
Samples are obtained either through the traditional melt-quench process, or using the generative models.
All structural properties reported in the main text (e.g., Fig. \ref{fig:validation},\ref{fig:conditional}) are reported as either averages over these six samples (e.g., structural features in Fig. \ref{fig:validation}a--c) or distributions (e.g., mechanical properties in Fig. \ref{fig:validation}d).
This allows us to perform statistically relevant comparisons between generated and simulated structures by accounting for the distribution of structures and properties in amorphous materials.

\noindent\textbf{Pair distribution functions:}
short-range structural ordering was characterized through pair distribution function (PDF) analysis.
All PDFs were calculated using OVITO\cite{ovito} with 100 bins between 0 and 8 \AA.

\noindent\textbf{Ring size distribution:}
the medium-range order structure of amorphous silica was characterized through ring size distribution analysis. These calculations were performed using the RINGS package,\cite{le2010ring} with Si–O bonds defined using a 2.0 \AA{} cutoff distance for ring identification within the network structure, which is in agreement with the partial PDF for Si-O pairs in Fig. \ref{fig:validation}a.

\begin{figure}[!h]
    \centering
    \includegraphics[width=1.0\linewidth]{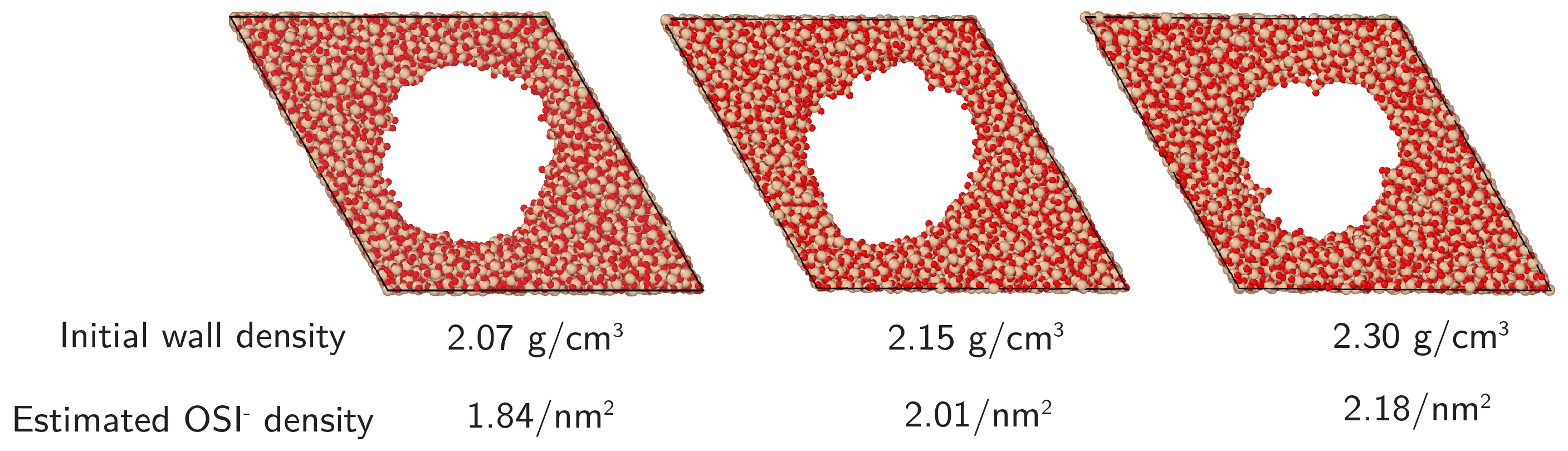}
    \caption{Mesoporous a-SiO2 structures generated by the denoiser model at varying initial wall densities.
    All structures are generated from the same initial geometry and processing conditions.
    Higher wall densities result in correspondingly denser non-bonding oxygen groups on pore surface.
    }
    \label{fig:si:pore-sio2}
\end{figure}

\noindent\textbf{Concentration of non-bridging oxygens:}
we quantified the concentration of non-bridging oxygens (NBOs) on generated mesoporous silica surfaces by calculating the ratio of NBO atoms to the estimated surface area of the pore.
As described in the main text, the pore morphology exhibits density-dependent shape and distortions from ideal cylindrical structures, as illustrated in Fig. \ref{fig:si:pore-sio2}.
To address this non-ideal geometry, we analyzed the radial distribution of Si and O atoms relative to the central simulation axis.
The effective pore radius was estimated as the distance at which the density of Si and O atoms plateaus, indicating the boundary between pore space and silica framework.
Whereas our generation has been performed only for the SiO$_2$ composition, NBOs are of high relevance to applications of mesoporous silica, where they can become silanol groups in the presence of hydrogen.

\noindent\textbf{Atomic volume:}
We employed Voronoi analysis in OVITO\cite{ovito} to compute the atomic volumes of CuZr structures.
Particle radii of 1.35 \r{A} for Cu and 1.55 \r{A} for Zr atoms are applied for analysis, with relative face area threshold set to 1\%.

\noindent\textbf{Voronoi indices:}
Voronoi indices for CuZr structures were computed using OVITO\cite{ovito} for polydisperse Voronoi tessellation with particle radii of 1.35 \r{A} for Cu and 1.55 \r{A} for Zr atoms.
Surface atoms occupying less than 1\% of the total surface area were excluded to minimize effect from experimental measurement and structural reconstruction errors.
Voronoi indices are represented using reduced Schlaefli notation, $\langle n_3,n_4,n_5,n_6 \rangle$, where $n_i$ denotes the number of polyhedral faces containing $i$ edges.
The polyhedral face distribution quantifies the frequency of faces with varying edge numbers.

We adopt the analysis functions published by Yang \textit{et al.}\cite{yang2021Determininga} to analyze Voronoi tessellation of atomic structures for both experimental and generated nanoparticles for consistency with that work.
Surface atoms occupying less than 1\% of the total surface area were excluded to minimize effect from experimental measurement and structural reconstruction errors.

\subsection*{Information entropy and QUESTS method} \label{sec:quests}

\noindent\textbf{Representation:} the representation of atomic environments was computed as described in our previous work.\cite{schwalbekoda2025information}
In summary, a number of $k = 32$ neighbors was used to represent the atomic environment, with a cutoff of 5 \AA.
No distinction was made between element types, which is implicitly recovered based on the coordination environments, and thus captured in the information entropy of the system.

\noindent\textbf{Information entropy:} the information entropy of descriptor distributions was computed as described before.\cite{schwalbekoda2025information}
Given a set of feature vectors $\Xset$, their information entropy is computed as follows:

\begin{equation}\label{eq:entropy}
    \mathcal{H} (\mathbf{\{X\}}) = -\frac{1}{n}\sum_{i=1}^{n} \text{log} \Biggl[ \frac{1}{n}\sum_{j=1}^{n}K_h(\mathbf{X}_i, \mathbf{X}_j)\Biggr],
\end{equation}

\noindent where our choice of the kernel $K_h$ is the Gaussian kernel,

\begin{equation}
    K_h(\mathbf{X}_i, \mathbf{X}_j) = \text{exp}\Biggl(\frac{-||\mathbf{X}_i - \mathbf{X}_j||^2}{2h^2} \Biggr),
\end{equation}

\noindent with a constant bandwidth $h = 0.015$ \AA$^{-1}$, as studied before.\cite{schwalbekoda2025information}

We define the differential entropy $\dH$ of a data point $\Y$ with respect to a reference dataset $\Xset$ as

\begin{equation}
    \delta\mathcal{H} (\text{Y} | \mathbf{\{X\}}) = - \text{log} \Biggl[ \sum_{i=1}^{n}K_h(\mathbf{Y}, \mathbf{X}_i)\Biggr].
\end{equation}

\noindent\textbf{Overlap:} From the definition of $\delta\mathcal{H}$, the overlap between two discrete distributions of atomic environments $\Yset$ and $\Xset$ is the fraction of environments $\Y_i \in \Yset$ with $\dH (\Y_i | \Xset) \leq 0$.
Therefore, a zero overlap between two distributions indicates that the distributions have disjoint supports, whereas a 100\% overlap indicates that the two distributions have the same support, even when the point-wise probabilities are different.
Within the context of generative models described in the main text, a high overlap indicates that the generated structures follow the same distribution of atomic environments, while a low overlap suggests that the structures do not share the same environments.

\noindent\textbf{Units:} Throughout this work, the natural logarithm was used for the entropy in Eq. \eqref{eq:entropy}, which scales the information to natural units (nats).

\section*{Data Availability}

The datasets generated in this work will be released upon publication of the manuscript.
The datasets of metallic glasses were obtained from the original source\cite{wang2020cuzr} at:
\url{https://figshare.com/articles/dataset/Heterogeneous_thermal_activation_energy_in_Cu-Zr_metallic_glasses/12485795}.
The datasets of the experimental nanoparticle was obtained from the original source\cite{yang2021Determininga} at:
\url{https://github.com/AET-MetallicGlass/Supplementary-Data-Codes}

\section*{Code Availability}

The code from this work was based on the code from Hsu \textit{et al.},\cite{Hsu2024Score} available at \url{https://github.com/LLNL/graphite}.
The code for computing Voronoi indices of amorphous nanoparticles was from Yang \textit{et al.},\cite{yang2021Determininga} available at \url{https://github.com/AET-MetallicGlass/Supplementary-Data-Codes}.
The information theoretical analysis was performed with the QUESTS package from Schwalbe-Koda \textit{et al.},\cite{schwalbekoda2025information} available at \url{https://github.com/dskoda/quests}.
The codebase to reproduce the results/plots from this work will be released at publication time.

\section*{Acknowledgements}

This work was supported by Toyota Research Institute under the Synthesis Advanced Research Challenge.
This work used computational and storage services associated with the Hoffman2 Shared Cluster provided by UCLA Office of Advanced Research Computing’s Research Technology Group.
The authors thank Linda Hung, Amalie Trewartha, Steven Torrisi, Fei Zhou, Jiawei Guo, and Long Qi for discussions and suggestions regarding this work.
The authors also thank Jianwei (John) Miao and his group for making the experimental dataset available.

\section*{Conflicts of Interest}

The authors have no conflicts to disclose.

\section*{Author Contributions}

\noindent\textbf{Kai Yang:} Formal Analysis; Data Curation; Investigation; Software; Validation; Visualization; Writing - Original Draft; Writing - Review \& Editing.
\noindent\textbf{Daniel Schwalbe-Koda:} Conceptualization; Formal Analysis; Investigation; Methodology; Project Administration;  Visualization; Writing - Original Draft; Writing - Review \& Editing; Funding Acquisition; Supervision.



\clearpage
\beginsupplement

\beginsuppinfo
\customlabel{sec:sinfo}{Supplementary Information}

\section{Supplementary Text}
\customlabel{sec:stext}{Supplementary Text}

\subsection{Evaluating generative models for amorphous materials}
\label{sec:si:evaluating}

Good generative models for amorphous materials should be able to reproduce quantitative distributions of structural features and properties for these systems.
Traditional validation strategies in amorphous materials rely on coordination environments, order parameters, energy distributions, or similar structural features that can be easily computed from the structure.
Deep learning methods often fall short of producing physically reasonable amorphous structures compared to simulation results --- exhibiting unphysical coordination environments, large deviations from expected pair correlation functions, or unrealistic defects --- let alone reproducing their properties or accounting for processing parameters such as cooling rates.\citeSupp{yong2024dismai-SI}
As examples of recent work related to amorphous materials, Yang \textit{et al.} presented a forward predictive model and an inverse generative approach for liquid electrolyte compositions,\citeSupp{yang2025unified-SI} though not all tasks are performed at the atomistic scale.
Recently, Chen \textit{et al.} used a graph variational autoencoder, with energy and radial distribution function (RDF) regularizations, to generate glass structures but only validate their generations with energy distributions and RDFs.\citeSupp{chen2025physical-SI}
Furthermore, qualitative visualizations or distance/descriptor distributions are insufficient to prove whether generated structures are simultaneously accurate, novel, and valid.
The work from Yong \textit{et al.}\citeSupp{yong2024dismai-SI} thoroughly benchmarks recent generative models' ability to produce realistic samples of amorphous and disordered systems, highlighting that most models cannot produce amorphous structures even at the qualitative level. In that work, structural characteristics and properties relevant to the analysis of amorphous materials have not been explored, however.
Kwon \textit{et al.} showed that a generative diffusion model can be used to generate amorphous carbon matching conditioned on characterization data.\citeSupp{kwon2024Spectroscopy-guided-SI}
Nevertheless, the resulting structures still showed outlier environments, and it remained unclear whether generative models can be used to generate and reproduce the properties of more complex amorphous materials such as multi-component ones, especially when conditioned to processing parameters.

In this work, we proposed a suite of tests to validate the quality of generated amorphous structures.
As a first figure of merit, structures are evaluated according to their short- and medium-range order, which includes pair distribution functions, bond angles, and network topology.
These distributions are commonly used to understand and analyze the structure of amorphous materials, and often correlated with experimental observations.
As a second tier of tests, we computed the macroscopic properties of simulated and generated structures such as their elastic tensor and plastic behavior.
These mechanical properties are highly sensitive to outlier environments, as minor unphysical features can produce large variations in energy (see Sec. \ref{sec:si:sensitivity} below), thus providing an excellent test case to assess the robustness of the generative models.
Finally, we introduced an information-theoretical metric to evaluate whether structures are sampled from similar probability distribution of atomistic environments, which is useful to quantify novelty (see Sec. \ref{sec:si:novelty} below).
These strategies are summarized in Fig. \ref{fig:overview}d of the main text.

\subsection{Sensitivity of macroscopic properties to outliers}
\label{sec:si:sensitivity}

While structures generated with the denoising process without extra noise can capture peak positions and demonstrate reasonable agreement with simulated partial PDFs (Fig. \ref{fig:si:01-PDF-comparison}a), they fail to reproduce the macroscopic properties of amorphous silica.
Macroscopic properties such as mechanical properties and stress-strain behavior are quite sensitive to structural fidelity and are thus excellent tests to the quality of generated structures.
Nonphysical atomic environments resulting from imperfect denoising can produce unrealistic bond lengths and angles that remain nearly undetected at short- and medium-range order (PDF, bond angle distributions, and ring size distributions such as Fig. \ref{fig:validation} of the main text).
However, even a single structurally invalid atom can result in major deviations in macroscopic properties, severely limiting the utility of generated structures for subsequent investigations.

\begin{figure}[!h]
    \centering
    \includegraphics[width=1.0\linewidth]{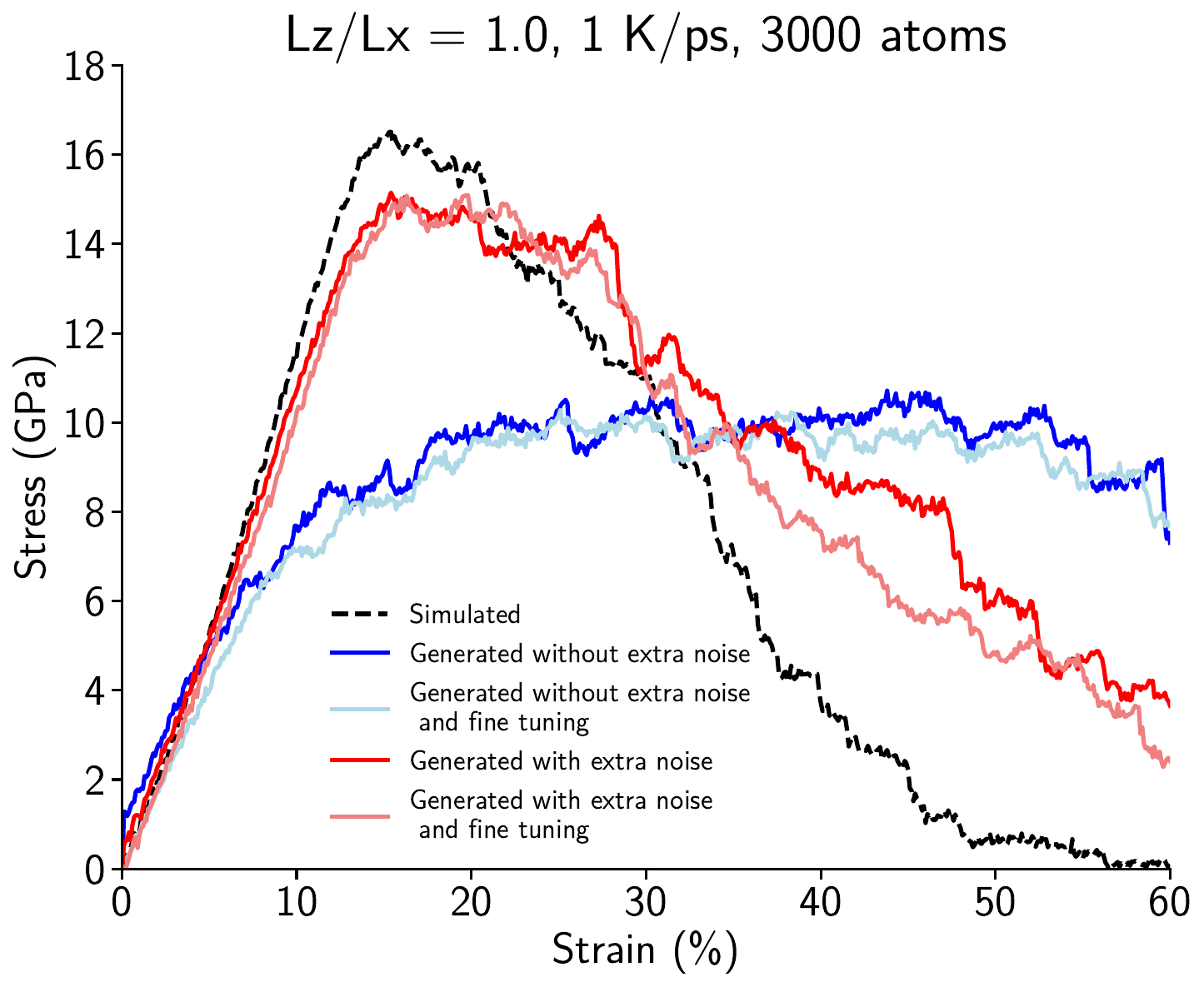}
    \caption{Stress-strain curves of simulated a-SiO$_2$ structure (dashed line) and generated structures (solid lines).
    Generated structures with extra noise and fine tuning show the best agreement compared to simulated results in terms of strength and ductility.
    Generated structures with extra noise show reasonable agreement compared to simulated result, but overestimate the stress at low strain and show more ductility.
    Generated structures without extra noise, even with fine tuning process, fail to reproduce the stress-strain behavior, showing a complete ductile behavior and underestimate the ultimate strength.
    All atomic structures were cube ($L_z / L_x$ = 1) with 3000 atoms, and generated with a target cooling rate of 1 K/ps.
    }
    \label{fig:si:fracture-fidelity}
\end{figure}

Figure \ref{fig:si:fracture-fidelity} compares stress-strain curves for structures generated under identical conditions using four methods: without extra noise, with extra noise, and each condition followed by a short MD equilibration (refinement).
Despite very similar partial PDF performance across all samples (Fig. \ref{fig:si:01-PDF-comparison}), their mechanical properties differ dramatically.
Structures generated without extra noise fail to exhibit realistic fracture behavior and yield strength compared to simulated references even after a post-generation MD equilibration to reduce the occurence of outliers.
Adding substantial extra noise during the denoising stage produces much more reasonable stress-strain curves, with brittle fracture characteristics.
Both non-refined approaches exhibit residual stress due to fixed density constraints during denoising.
In contrast, structures generated with extra noise and MD refinement result in similar stress-strain curves given the expected variability of ductility across different samples, reproducing both yield strength and elastic modulus.

Quantitative analysis of the elastic properties further demonstrates this sensitivity.
For all a-SiO$_2$ structures, we computed the average bulk modulus ($K$), shear modulus ($G$), Young's modulus ($E$), and Poisson's ratio ($\nu$) across simulated and generated results.
Table \ref{tab:si:elastic} summarizes these results.
For the reference simulated structures, we obtain $K = 61.2$ GPa, $G = 37.1$ GPa, $E = 92.4$ GPa, and $\nu = 0.247$, compatible with the mechanical properties for this interatomic potential.\citeSupp{yang2019predicting-SI}
Generated structures without extra noise, despite exhibiting somewhat reasonable PDFs (Fig. \ref{fig:si:01-PDF-comparison}), produce severely aberrant values: $K = 166.7$ GPa, $G = 2.9$ GPa, $E = 8.7$ GPa, and $\nu = 0.491$.
The post-generation MD refinement corrects the elastic properties of generated structures, resulting in values of $K = 63.2$ GPa, $G = 37.5$ GPa, $E = 93.9$ GPa, and $\nu = 0.256$.
Interestingly, although these values are close to simulated reference, the failure of exhibiting fracture behavior (Fig. \ref{fig:si:fracture-fidelity}) demonstrates that multiple validation approaches are essential for assessing generated structure fidelity, and underscores the critical importance of adding extra noise during the generation process.
Adding extra noise during denoising significantly improves the properties of generated structures, but still exist small discrepancies, leading to $K = 67.0$ GPa, $G = 32.6$ GPa, $E = 83.6$ GPa, $\nu = 0.287$.
Finally, after a short MD equilibration, structures generated with extra noise achieve excellent agreement with simulated results: $K = 60.6$ GPa, $G = 37.8$ GPa, $E = 93.8$ GPa, and $\nu = 0.241$.
These findings underscore that mechanical properties provide a stringent test of structural fidelity, revealing deficiencies that remain hidden in conventional structural analyses.

\begin{table}[H]
\centering
\caption{
Average mechanical properties (bulk modulus $K$, shear modulus $G$, Young's modulus $E$, and Poisson's ratio $\nu$) of a-SiO$_2$ structures for different structure generation methods.
All values are in GPa except for $\nu$ (dimensionless).
Errors correspond to the standard deviation of properties across six generated and simulated samples.
Samples generated with extra noise and MD refinement exhibit the exact distribution of elastic properties of the simulated reference configurations.
}
\begin{tabular}{lcccc}
\hline
Method & $K$ (GPa) & $G$ (GPa) & $E$ (GPa) & $\nu$ \\
\hline
Generated (no extra noise) & 166.7 $\pm$ 8.2 & 2.9 $\pm$ 0.3 & 8.7 $\pm$ 1.1 & 0.491 $\pm$ 0.001 \\
Generated (no extra noise) + MD refinement & 63.2 $\pm$ 2.1 & 37.5 $\pm$ 2.3 & 93.9 $\pm$ 4.5 & 0.256 $\pm$ 0.016 \\
Generated (with extra noise) & 67.0 $\pm$ 8.8 & 32.6 $\pm$ 4.3 & 83.6 $\pm$ 8.7 & 0.287 $\pm$ 0.042 \\
Generated (with extra noise) + MD refinement & 60.6 $\pm$ 3.1 & 37.8 $\pm$ 0.8 & 93.8 $\pm$ 1.4 & 0.241 $\pm$ 0.014\\ \hline
Simulated reference & 61.2 $\pm$ 2.9 & 37.1 $\pm$ 3.0 & 92.4 $\pm$ 5.6 & 0.247 $\pm$ 0.025 \\
\hline
\end{tabular}
\label{tab:si:elastic}
\end{table}

\subsection{Overlap as novelty metric of amorphous structures}
\label{sec:si:novelty}

In the main text, we introduced the use of information theoretical quantities to measure the ``novelty'' of newly sampled amorphous configurations.
In generative models for molecules, for instance, novel data points are molecular graphs distinct from the training set,\citeSupp{gomez2018automatic-SI,brown2019guacamol-SI} and thus can be trivially computed by comparing the molecular graphs.
In generative models for inorganic crystalline materials, novelty is defined by comparing the space groups, atomic basis, compositions, and other discrete quantities that uniquely define a crystal.\citeSupp{xie2022crystal-SI,zeni2025generative-SI}
Whereas this is not as trivial as in the case of molecules and can lead to deviations in the assignment of ``novelty'' of a crystal,\citeSupp{leeman2024challenges-SI} it still allows for quantitative comparisons of novelty in crystal structure generation.

\begin{figure}[!h]
    \centering
    \includegraphics[width=1.0\linewidth]{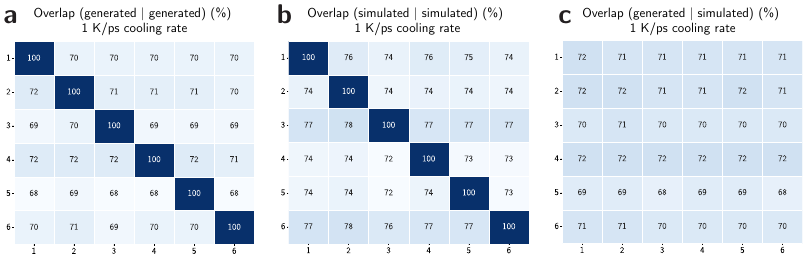}
    \caption{Overlap score matrices for \textbf{a}, generated $|$ generated, \textbf{b}, simulated $|$ simulated, and \textbf{c}, generated $|$ simulated a-SiO$_2$ structures.
    Each matrix computes the overlap scores between two series of six independent simulations or generations.
    High overlap values indicate strong structural similarities.
    The results show two main results: (1) the distribution of generated atomic environments is very similar to the ones obtained from MD simulations, showing that the model produces valid amorphous structures; and (2) the generative model does not simply ``copy-and-paste'' the known amorphous structures, and instead produces novel configurations, with a novelty rate nearly identical to the ones obtained from simulations.
    All structures are prepared with 3000 atoms and under 1 K/ps cooling rate.
    }
    \label{fig:si:overlap}
\end{figure}

However, in contrast to crystalline structures or molecular graphs, novelty is challenging to define in the amorphous space.
Even amorphous materials obtained from identical simulation conditions will not be identical unless they use exactly the same initial configuration and random seeds, leading to an ill-defined concept of ``novelty'' in the amorphous materials space.
Thus, analyzing whether a generative model can approximate the data distribution and produce new samples requires measuring probabilities in the data domain.
To do that, we quantified the similarity between amorphous configurations using an information-theoretical method.
Specifically, we computed a probability distribution over environments for a-SiO$_2$ structures following our previous work,\citeSupp{schwalbekoda2025information-SI} and evaluated the similarity between distributions using an information-theoretical overlap score.
This comparison between distributions resembles the Fr\'echet Inception Distance in generative adversarial networks,\citeSupp{heusel2017gans-SI} which accounts for novelty in terms of distribution statistics rather than point-wise metrics.
As discussed in the main text, we propose that the overlap of distributions of atomic environments can provide insights on the ``novelty'' of amorphous structures.
These overlaps are computed per pair of structures, and thus can be depicted as a non-symmetrical matrix, as $P(A | B) \neq P(B | A)$.
Figure \ref{fig:si:overlap} shows how these matrices look like when six a-SiO$_2$ structures are compared against other six structures.
The diagonal of Figs. \ref{fig:si:overlap}a,b indicate that the overlap of a structure with itself is 100\%, as expected.
Off-diagonal elements show that simulated a-SiO$_2$ samples (``Sim $|$ Sim'') exhibited, on average, 75\% overlap with each other despite deriving from the same processing condition, with narrow variability between the overlaps (Fig. \ref{fig:si:overlap}b).
This result shows that, for 3000-atom systems sampled in our simulations, about 75\% of the environments are consistently sampled, and 25\% environments are usually considered ``new'' even among deterministic simulations.
Similarly, generated samples showed ~70\% overlap with each other (``Gen $|$ Gen'' in Fig. \ref{fig:validation}e).
This overlap indicates that the generative model is capturing both the information relevant to describe an amorphous state and the approximate expected novelty in the amorphous states.

To test for mode collapse, we also computed the overlap between the distribution of generated environments and simulated ones (``Gen $|$ Sim''), shown in Fig. \ref{fig:si:overlap}c.
If the model displayed a mode collapse and only learned to reproduce the training data, this overlap would be close to 100\%.
On the other hand, a model that consistently produces unphysical data points would exhibit very low overlap with the simulated structures even when the ``Gen $|$ Gen'' overlap is high.
Finally, a model that perfectly reproduces the distribution of simulated data should exhibit an overlap close to 75\%.
In our case, the overlap between the generated and simulated structures was approximately 70\%, thus demonstrating that our generated a-SiO$_2$ samples maintain excellent structural fidelity while preserving appropriate configurational diversity.
Figure \ref{fig:validation}e of the main text summarizes these findings by comparing the distribution of overlaps.

\subsection{Importance of an extra noise in denoising process}
\label{sec:si:noise}

Adding a high level of extra noise during the denoising process is critical to generate valid and novel amorphous structures.
The noise magnitude $\sigma$ is analogous to a temperature in conventional melt-quench molecular dynamics simulations, a connection that has been explored before in the context of Langevin dynamics within diffusion models.\citeSupp{ho2020denoising-SI}
High $\sigma$ values in the beginning of the denoising process are analogous to elevated temperatures, where atoms possess sufficient kinetic energy to escape local minima and break structural correlations imposed by the initial state.
Furthermore, the progressive reduction of $\sigma$ to zero during denoising is similar to the controlled cooling process in MD quenching.
The only difference in this analogy is that the generative model can produce structures with different cooling rates at a fixed number of steps, whereas an MD-based melt-quench process requires long simulations to properly cool the structures.
Denoising without extra noise resembles instantaneous quenching from a liquid to a glassy state, preventing adequate structural relaxation.

\begin{figure}[!h]
    \centering
    \includegraphics[width=1.0\linewidth]{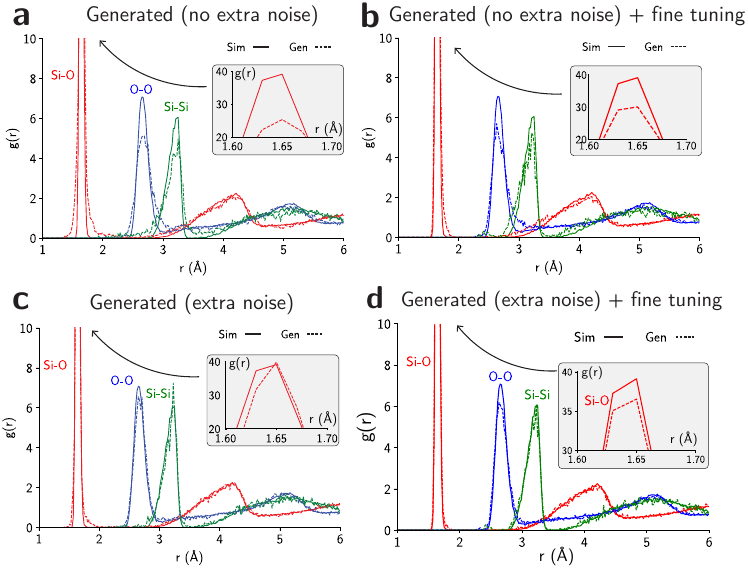}
    \caption{Comparison of partial pair distribution functions $g(r)$ between simulated and generated a-SiO$_2$ for cases: \textbf{a}, generated without adding extra noise, \textbf{b}, generated without adding extra noise, followed by MD refinement, \textbf{c}, generated with extra noise, and \textbf{d}, generated with extra noise and MD refinement).
    All cases demonstrate that generated configurations reproduce good structures according to structural quality metrics.
    Adding extra noise and fine tuning with MD removes most nonphysical behaviors (such as short Si–O pairs) and give accurate peaks for PDFs without intensity decreases.
    }
    \label{fig:si:01-PDF-comparison}
\end{figure}

Partial pair distribution function (PDF) analysis in Fig. \ref{fig:si:01-PDF-comparison} shows that amorphous structures with superior structural quality are obtained when extra noise is employed during the denoising process.
Importantly, when approximating the distribution of atomistic environments in the training data, the generative model makes no consideration about the energy of the system.
Sometimes, this leads to a few outlier environments that, while rare, can be nonphysical for the simulated system, such as Si-O bonds shorter than 1.5 \AA~ in a-SiO$_2$ (Fig. \ref{fig:si:01-PDF-comparison}a,b).
While both approaches generate amorphous structures with reasonable short-range order, structures generated with extra noise exhibit denser bond pair distributions and more accurate peak intensities compared to reference simulations (Fig. \ref{fig:si:01-PDF-comparison}c,d).
Directly applying MD fine tuning process on structures generated without extra noise can eliminate unphysical short Si–O bonds and enhance partial PDF performance; however, the second coordination shell remains poorly reproduced (Fig. \ref{fig:si:01-PDF-comparison}b).
On the other hand, the small amounts of outlier environments in structures generated with extra noise can be readily corrected through picosecond-long MD simulations after the denoising process and show excellent agreement with the simulated reference  (Fig. \ref{fig:si:01-PDF-comparison}d).

\begin{figure}[!h]
    \centering
    \includegraphics[width=1.0\linewidth]{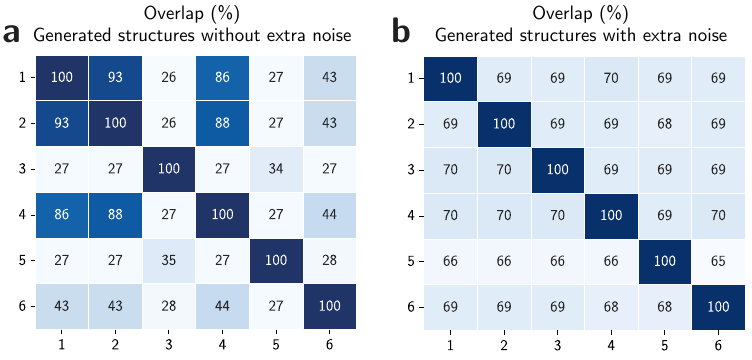}
    \caption{
    Overlap score matrices for generated a-SiO$_2$ structures \textbf{a}, without and \textbf{b}, with extra noise during the diffusion model.
    High overlap values indicate structural similarity, while low values indicate diversity.
    Without extra noise, structures show variable similarity.
    For instance, samples 1, 2, and 4 generated without noise are overly similar to each other, indicating a mode collapse into a single configuration.
    On the other hand, others differ substantially, indicating that the right distribution of environments is not captured.
    When extra noise is added, structures exhibit more consistent similarity across all samples.
    }
    \label{fig:si:01-overlap-noise-comparison}
\end{figure}

\begin{figure}[!h]
    \centering
    \includegraphics[width=1.0\linewidth]{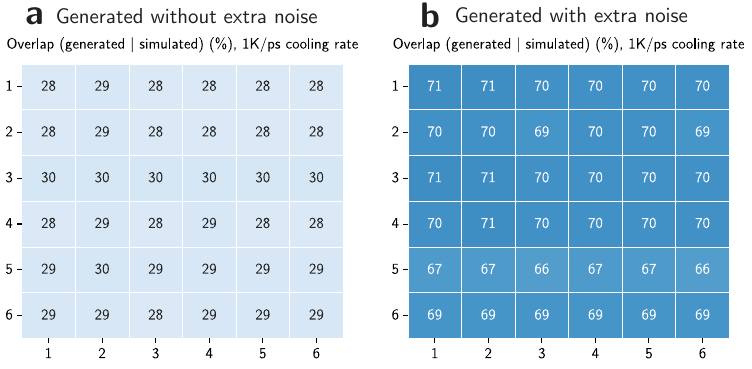}
    \caption{Overlap score matrices for generated against simulated a-SiO$_2$ structures \textbf{a}, without and \textbf{b}, with extra noise during generation.
    High overlap values indicate that the distribution of environments is similar.
    Generated samples without extra noise are mostly dissimilar to the simulated samples.
    On the other hand, the samples generated with extra noise are nearly as similar to simulated ones as simulated samples are to themselves.
    }
    \label{fig:si:01-overlap-noise-to-simu-comparison}
\end{figure}

As discussed in the main text and Methods, structural diversity of amorphous structures was quantified using overlap scores between six independently generated a-SiO$_2$ structures.
Without extra noise, overlap scores were distributed either excessively high (approaching 100\%, indicating mode collapse) or unusually low (below 45\%, suggesting structural inaccuracy) as shown in Fig. \ref{fig:si:01-overlap-noise-comparison}a.
Conversely, structures generated with extra noise exhibited consistent overlap scores around 70\%, matching nearly perfectly the balance between diversity and validity observed in simulated structures (Fig. \ref{fig:si:01-overlap-noise-comparison}b).
When compared to MD simulated reference structures, configurations generated without extra noise showed poor correlation (overlap scores around 29\%, Fig. \ref{fig:si:01-overlap-noise-to-simu-comparison}a), whereas those with extra noise maintained consistent accuracy with overlap scores around 70\% (Fig. \ref{fig:si:01-overlap-noise-to-simu-comparison}b).
Altogether, these figures of merit (PDF analysis, overlap score distributions, and structural diversity metrics) provide evidence on the importance of using a large extra noise when generating amorphous materials using denoiser models.

\subsection{Information theoretical analysis of conditionally generated silica glasses}
\label{sec:si:cooling-rate}

Given the known relationships between free volume, configurational entropy, and cooling rates in glasses,\citeSupp{berthier2019configurational-SI,wu2017thermodynamic-SI} we used our information entropy approach as a surrogate of configurational entropy for glassy materials.\citeSupp{schwalbekoda2025information-SI}
As explained in Section \ref{sec:si:novelty}, this method computes an information entropy of distributions of atomic environments and thus translates atomistic systems into probability distributions.
Within glassy materials, even at constant density, input cooling rates should be able to steer the final generated structure to different states containing different distributions of atomistic environments, and thus different values of configurational entropy.\citeSupp{gedeon2021configurational-SI}
In particular, rapid cooling produces high-energy states with incomplete structural relaxation, leading to glasses with more diverse local atomic motifs compared to slowly cooled counterparts, which is quantified by a higher information entropy.
The relationship between information entropy averaged across six generated structures, cooling rate, and density of generated structures is shown in Figure \ref{fig:conditional}a of the main text.
The results confirm that higher information entropy is achieved for higher cooling rates, even at constant densities, validating the model's ability to produce conditional generation of cooling rates.
For generated a-SiO$_2$ samples at the right densities (Fig. \ref{fig:si:02-entripy-coolingrate-comparison}), average information entropy values were 6.87, 6.70, 6.53, and 6.43 nats for cooling rates of 10$^{2}$, 10$^{1}$, 10$^{0}$, and 10$^{-1}$ K/ps, respectively.
These results are in excellent agreement with the average information entropy values of 7.01, 6.76, 6.58, and 6.43  nats for the reference simulated structures.

\subsection{Effect of cooling rate on simulated and generated properties}
\label{sec:si:cooling}

\begin{figure}[!h]
    \centering
    \includegraphics[width=1.0\linewidth]{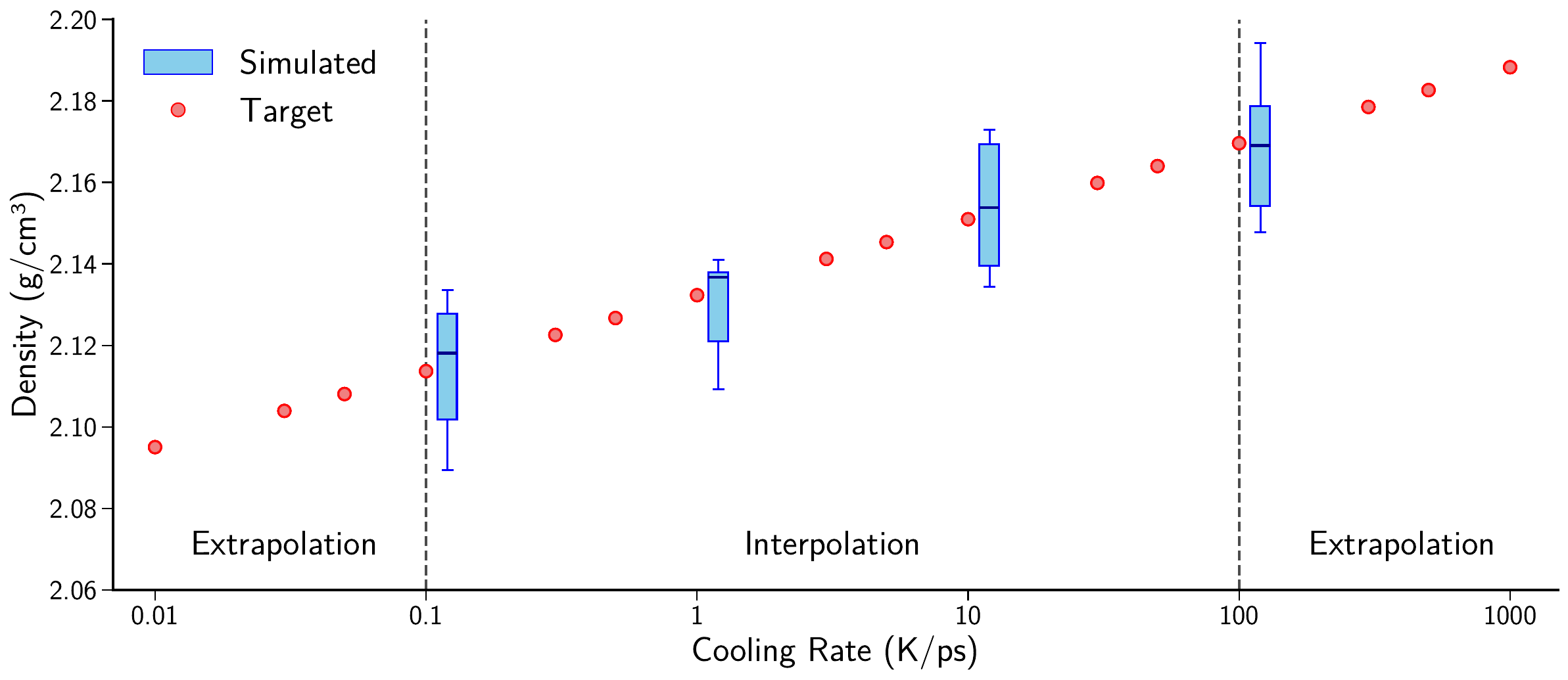}
    \caption{Densities of simulated a-SiO$_2$ structures and target densities across different cooling rates.
    Target densities are determined based on a linear regression model valid with simulated densities and extended to 0.01 and 1000 K/ps.
    }
    \label{fig:si:02-density-coolingrate-comparison}
\end{figure}

MD simulations are widely employed to study glass structure and properties, yet their inherent limitation to short timescales imposes unrealistically fast cooling rates on simulated structures.\citeSupp{li2017cooling-SI}
This constraint makes it challenging to provide a one-to-one comparison with experimental glasses processed under typical laboratory conditions.
In simulations of a-SiO$_2$, high cooling rates typically introduce spurious structural defects, though bulk structural characteristics remain relatively stable despite thermal history.
Conversely, properties such as density and thermal expansion coefficients exhibit strong dependence on cooling rate.
Since our denoising approach generates structures at fixed density (i.e., number of particles and volume), we developed a regression model that enables us to specify the density as a function of cooling rate (Fig. \ref{fig:si:02-density-coolingrate-comparison}).
Within the range of cooling rates of interest, simulated amorphous silica densities are reasonably linear with the logarithm of the cooling rate, simplifying the determination of initial densities for denoising.
Although the density of silica does not follow a linear trend across a wider range of cooling rates, \citeSupp{lane2015cooling-SI} they are valid within our study and simplify the analysis of model generalization.
The post-denoising MD equilibration employs both NVT and NPT ensembles, but the latter typically only increases the density by around 0.01 g/cm$^3$.

\begin{figure}[!h]
    \centering
    \includegraphics[width=1.0\linewidth]{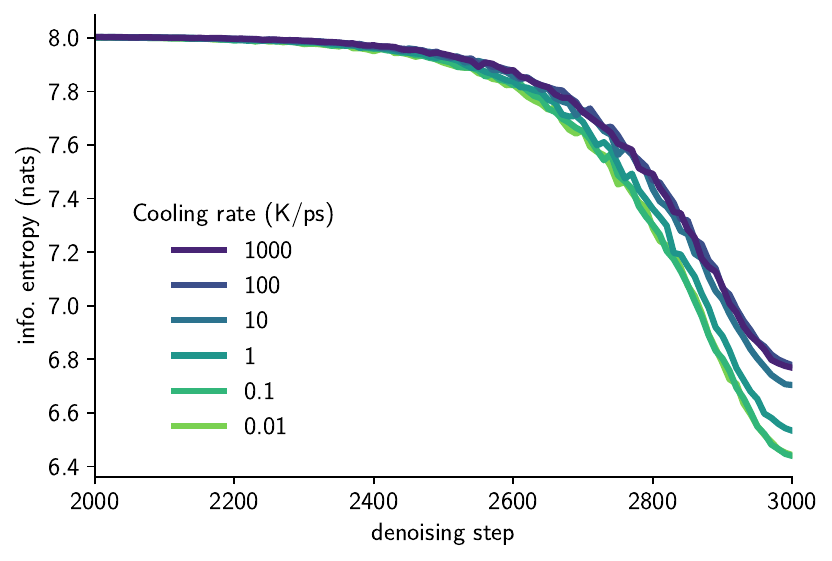}
    \caption{Information entropy of generated a-SiO$_2$ structures sampled along the denoising process with different target cooling rates.
    The denoising process was the same across the different conditions.
    The maximum information entropy that can be obtained for a system of this size with 3000 atoms is $\ln 3000 = 8.0$ nats.
    }
    \label{fig:si:02-entripy-coolingrate-comparison}
\end{figure}

Figure \ref{fig:si:02-entripy-coolingrate-comparison} illustrates how the information entropy changes at the final steps of conditional denoising across different cooling rates.
Following the methodology described in the Methods section, entropy values remain approximately 8.0 nats before step 2200, reflecting the maximum possible entropy with a system with 3000 atoms ($\ln 3000 = 8.0$).\citeSupp{schwalbekoda2025information-SI}
As the noise is progressively lowered due to the noise schedule, the information entropy of the system decreases and structures develop local ordering, with slower cooling rates leading to lower final entropy values, all consistent with simulated trends.

\begin{figure}[!h]
    \centering
    \includegraphics[width=1.0\linewidth]{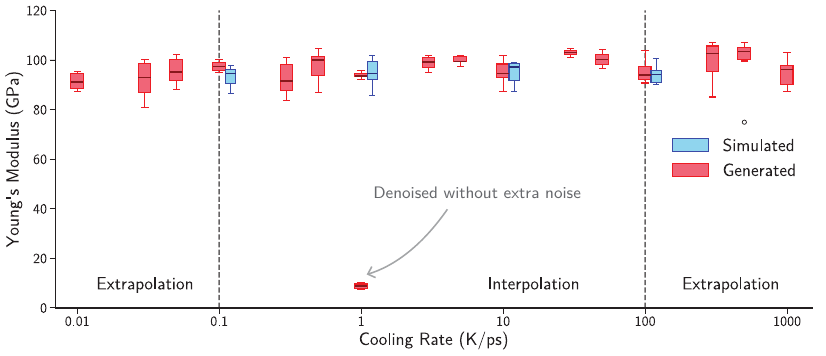}
    \caption{Young's modulus of simulated and generated a-SiO$_2$ samples across different cooling rates.
    The moduli are highly sensitive to outliers, yet the denoiser is able to generate structures with accurate Young’s moduli compared to simulations, even outside the training domain (10$^{-2}$ and 10$^3$ K/ps).
    The results of structures generated without adding extra noise demonstrate that high fidelity structures are necessary to reproduce reasonable elastic property.
    }
    \label{fig:si:young_cooling}
\end{figure}

\begin{figure}[!h]
    \centering
    \includegraphics[width=1.0\linewidth]{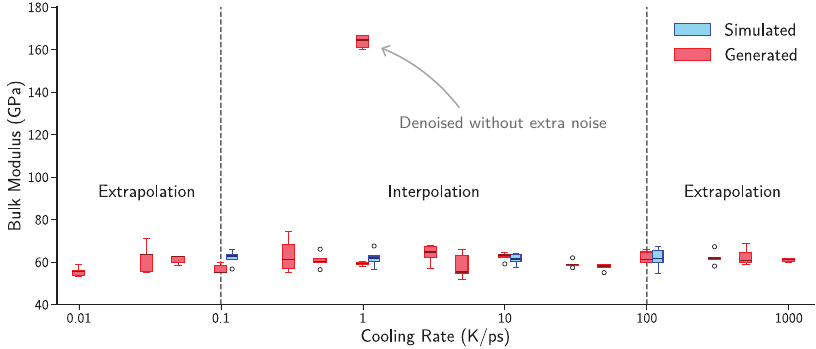}
    \caption{Bulk modulus of simulated and generated a-SiO$_2$ samples across different cooling rates.
    The moduli are highly sensitive to outliers, yet the denoiser is able to generate structures with accurate bulk moduli compared to simulations, even outside the training domain (10$^{-2}$ and 10$^3$ K/ps).
    The results of structures generated without adding extra noise demonstrate that high fidelity structures are necessary to reproduce elastic properties.
    }
    \label{fig:si:bulk_cooling}
\end{figure}

\begin{figure}[!h]
    \centering
    \includegraphics[width=1.0\linewidth]{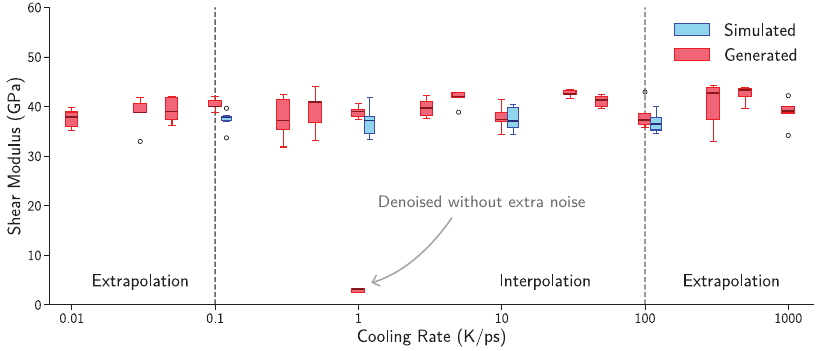}
    \caption{Shear modulus of simulated and generated a-SiO$_2$ samples across different cooling rates.
    The moduli are highly sensitive to outliers, yet the denoiser is able to generate structures with accurate shear moduli compared to simulations, even outside the training domain (10$^{-2}$ and 10$^3$ K/ps).
    The results of structures generated without adding extra noise demonstrate that high fidelity structures are necessary to reproduce elastic properties.
    }
    \label{fig:si:shear_cooling}
\end{figure}

\begin{figure}[!h]
    \centering
    \includegraphics[width=1.0\linewidth]{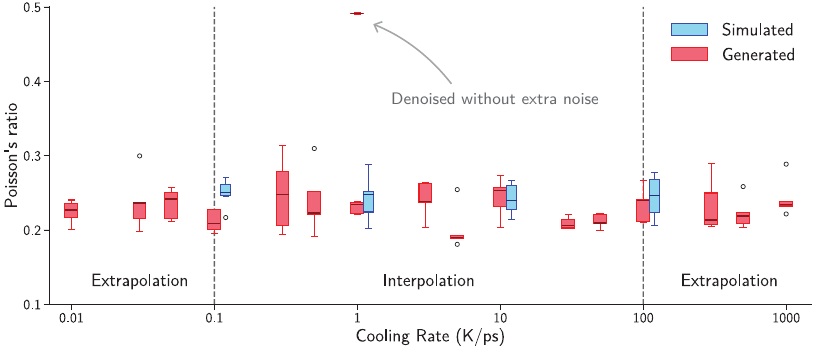}
    \caption{Poisson's ratio of simulated and generated a-SiO$_2$ samples across different cooling rates.
    The elastic properties are highly sensitive to outliers, yet the denoiser is able to generate structures with accurate Poisson's ratios compared to simulations, even outside the training domain (10$^{-2}$ and 10$^3$ K/ps).
    The results of structures generated without adding extra noise demonstrate that high fidelity structures are necessary to reproduce elastic properties.
    }
    \label{fig:si:poisson_cooling}
\end{figure}

Validation of structures as a function of cooling rate employs mean Si–O–Si bond angles and mechanical properties, including Young's modulus (Fig. \ref{fig:si:young_cooling}), bulk modulus (Fig. \ref{fig:si:bulk_cooling}), shear modulus (Fig. \ref{fig:si:shear_cooling}) and Poisson's ratio (Fig. \ref{fig:si:poisson_cooling}) across different cooling rates.
Generated structures accurately capture Si–O–Si bond angle trends with respect to cooling rate, demonstrating excellent agreement in both interpolation regions (0.3, 0.5, 3, 5, 30, and 50 K/ps) and ``generalization'' domains, i.e., targeted cooling rates outside of the interval of $[10^{-1}, 10^{2}]$ K/ps.
Mechanical properties, which exhibit high sensitivity to structural accuracy (see Sec. \ref{sec:si:sensitivity} above), remain physically consistent across all cooling rates for generated samples.
On the other hand, outliers, such as structures generated without extra noise (Fig. \ref{fig:si:01-PDF-comparison}a), exhibit aberrant elastic properties (Figs \ref{fig:si:young_cooling}--\ref{fig:si:poisson_cooling}), despite having similar partial PDFs compare to simulated samples.
This confirms that our conditioned structure generation produces structurally valid amorphous configurations across a range of cooling rates.

\subsection{Size and shape effects in fracture of amorphous silica}
\label{sec:si:fracture}

\begin{figure}[!h]
    \centering
    \includegraphics[width=1.0\linewidth]{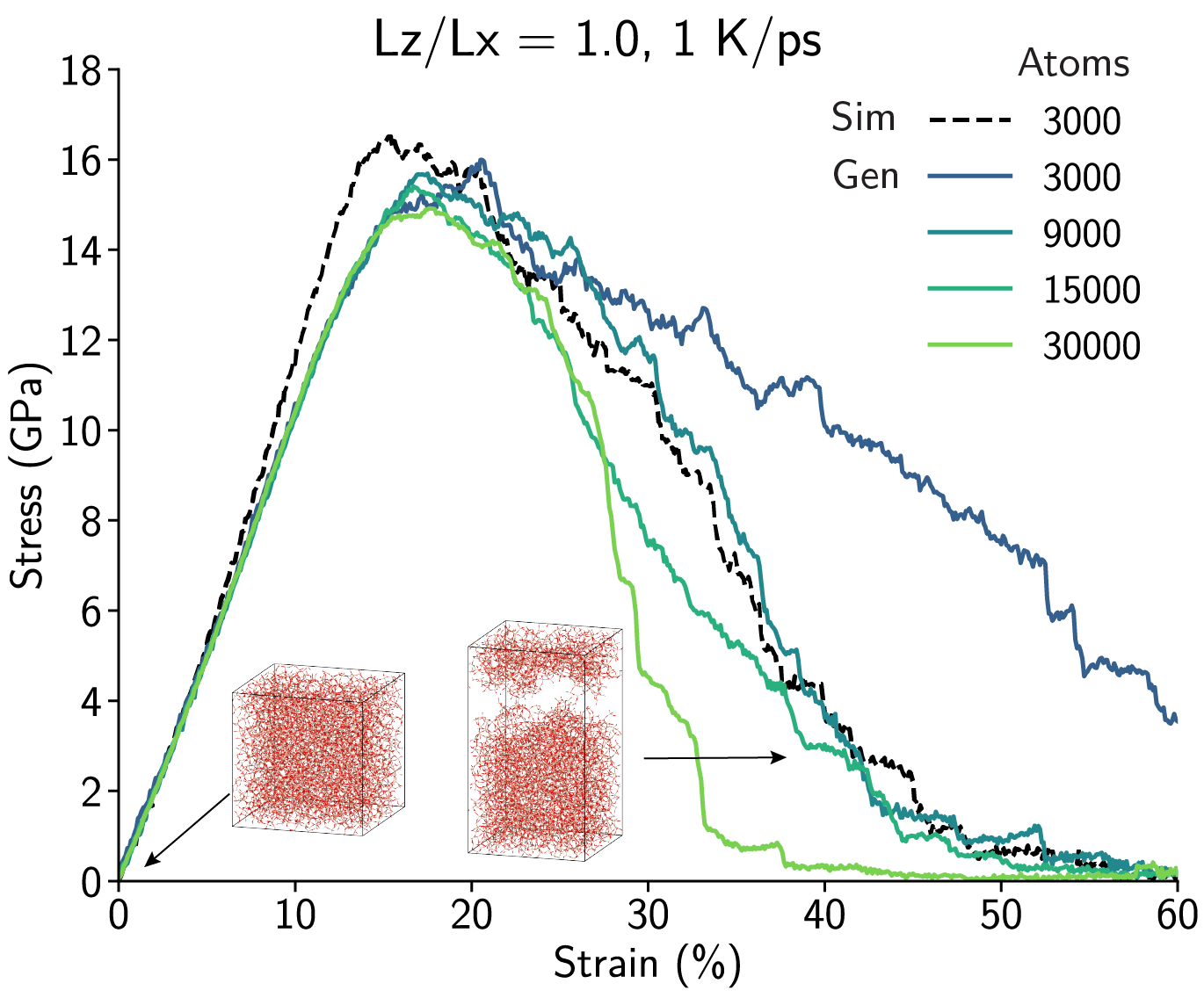}
    \caption{Stress-strain curves show that generated a-SiO$_2$ structures (solid lines) recover the expected trends of elastic and plastic deformation behavior across length scales, with more ductile fracture at smaller length scales.
    The stress-strain curve of a simulated a-SiO$_2$ structure with 3000 atoms is shown with a dashed line.
    The initial simulation box was a cube ($L_z / L_x$ = 1).
    }
    \label{fig:si:fracture_cube}
\end{figure}

Despite known variability in simulated plastic properties such as strength, generated structures consistently lead to stress-strain curves with good approximation of the elastic regime and ultimate tensile strength of the simulated material.
This is true even at large length scales, when cracks are strongly influenced by structural outliers.
Moreover, larger generated structures exhibit an increasingly brittle behavior, with substantial reduction in ductility from structures with 3,000 and 30,000 atoms, consistent with past experimental results.\cite{}
Figure \ref{fig:si:fracture_cube} shows the stress-strain curves, averaged from six independent tests, for the fracture of generated cubes (L$_z$/L$_x$ = 1.0) with system sizes ranging from 3,000 to 30,000 atoms.
Given the larger aspect ratio of the generated samples compared to cubic structures in Fig. \ref{fig:si:fracture_cube}, brittle fracture is observed at lower strains in Fig. \ref{fig:fracture}a.
This agrees with previous studies stating that the length along the deformation axis is the most important factor in minimize boundary effects and provide adequate volume for realistic crack propagation.\citeSupp{barciela2024size-SI}

\subsection{Voronoi indices of experimental nanoparticle model}
\label{sec:si:voronoi}

We adopt the functions from Yang \textit{et al.}\citeSupp{yang2021Determining-SI} to calculate the Voronoi indices distribution for both experimental and generated nanoparticle models.
In particular, we compare the top Voronoi indices in both experimental and generated structures.
As shown in Table \ref{tab:voronoi_comparison}, which contains the top-10 Voronoi indices in the experimental structures, our denoiser model successfully generates atomic structures that capture most of the Voronoi indices observed in the experimental model except for $\langle 0,4,5,3 \rangle$.
This indicates that the top Voronoi indices found in the experimental data are also the top Voronoi indices generated by the diffusion model.
Nevertheless, deviations are still found to be typically within 1-2\% for the majority of configurations.
Given the sensitivity of Voronoi indices in representing short-range ordering, they create a stringent figure of merit for evaluation of generative models for amorphous materials, and can be the focus of future investigations.

\begin{table}[htb!]
\centering
\caption{
Fraction of Voronoi indices of experimental and generated nanoparticle models. The top indices from both experimental and generated atomic model are computed and compared according to their frequency.
The top 10 Voronoi indices from experimental model and corresponding fractions from generated model are shown here.
There is a strong overlap between the top Voronoi indices for both the generated and experimentally determined structures, with only one of the Voronoi indices in the experimental model not present in the top indices of the generated counterparts.
The Voronoi indices follow a reduced Schlaefli notation, $\langle n_3,n_4,n_5,n_6 \rangle$, with $n_i$ indicating the number of polyhedron faces with $i$ edges/vertices.
}
\label{tab:voronoi_comparison}
\begin{tabular}{ccc}
\hline
\textbf{Voronoi Indices} & \textbf{Experimental (\%)} & \textbf{Generated (\%)} \\
\hline
$\langle 0,4,4,3 \rangle$ & 5.36 & 3.28 \\
$\langle 0,3,6,3 \rangle$ & 5.28 & 5.53 \\
$\langle 0,4,4,2 \rangle$ & 3.93 & 2.40 \\
$\langle 0,4,4,4 \rangle$ & 3.86 & 2.29 \\
$\langle 0,3,6,2 \rangle$ & 3.84 & 2.68 \\
$\langle 0,3,6,4 \rangle$ & 3.62 & 6.72 \\
$\langle 0,2,8,1 \rangle$ & 3.19 & 4.70 \\
$\langle 0,2,8,2 \rangle$ & 3.19 & 6.35 \\
$\langle 0,3,6,1 \rangle$ & 2.15 & 2.12 \\
$\langle 0,4,5,3 \rangle$ & 1.74 & Not in top 20 \\

\hline
\end{tabular}
\end{table}

\clearpage
\section{Supplementary Figures}

\clearpage

\end{document}